\documentclass[ejs]{imsart}

\usepackage{amssymb}
\usepackage{graphicx}

\RequirePackage[OT1]{fontenc}
\RequirePackage{amsthm,amsmath}
\RequirePackage{natbib}
\usepackage{multirow}
\usepackage{multicol}
\usepackage[usenames,dvipsnames]{xcolor}
\RequirePackage[colorlinks,citecolor=blue,urlcolor=blue]{hyperref}
\usepackage{breakcites}

\startlocaldefs
\numberwithin{equation}{section}
\theoremstyle{plain}

\endlocaldefs

\newcommand{\be}{\begin{equation}}
\newcommand{\ee}{\end{equation}}
\newtheorem{corollary}{Proposition}
\newtheorem{theorem}{Theorem}
\newtheorem{lemma}{Lemma}

\newtheorem{remark}{Remark}
\newcommand\scalemath[2]{\scalebox{#1}{\mbox{\ensuremath{\displaystyle #2}}}}

\def\Ri{T_j}
\def\Yti{Y_{1j}}
\def\Yci{Y_{0j}}
\def\bx{\boldsymbol{x}_j} 
\def\bfx{\boldsymbol{x}} 
\def\bst{\mathcal{S}_1}
\def\bsc{\mathcal{S}_0}
\def\bs{\mathcal{S}}
\def\htau{\hat{\tau}_j}
\def\bz{\tilde{\boldsymbol{x}}_j}

\def\mutIPW{\hat{\mu}_{1\mbox{\tiny IPW1}}}
\def\mucIPW{\hat{\mu}_{0\mbox{\tiny IPW1}}}
\def\thetaIPW{\hat{\theta}_{\mbox{\tiny IPW1}}}
\def\mutH{\hat{\mu}_{1\mbox{\tiny IPW2}}}
\def\mucH{\hat{\mu}_{0\mbox{\tiny IPW2}}}
\def\thetaH{\hat{\theta}_{\mbox{\tiny IPW2}}}
\def\bPsi{\boldsymbol{\Psi}}

\def\sbstj{\sum_{j\in \bst}}
\def\sbscj{\sum_{j\in \bsc}}
\def\sbsj{\sum_{j\in \bs}}
\def\sbsji{\sum_{j\in \bs_i}}

\def\mutDRO{\hat{\mu}_{1\mbox{\tiny AIPW1}}}
\def\mucDRO{\hat{\mu}_{0\mbox{\tiny AIPW1}}}
\def\thetaDRO{\hat{\theta}_{\mbox{\tiny AIPW1}}}
\def\mutDRT{\hat{\mu}_{1\mbox{\tiny AIPW2}}}
\def\mucDRT{\hat{\mu}_{0\mbox{\tiny AIPW2}}}
\def\thetaDRT{\hat{\theta}_{\mbox{\tiny AIPW2}}}
\def\thetaDR{\hat{\theta}_{\mbox{\tiny DR}}}

\def\thetaP{\hat{\theta}_{\mbox{\tiny PEL}}}
\def\thetaPC{\hat{\theta}_{\mbox{\tiny MCP}}}
\def\thetaPCB{\hat{\theta}_{\mbox{\tiny MCP}}^{[b]}}

\def\lambdaPC{\hat{\boldsymbol{\lambda}}_{\mbox{\tiny MCP}}}
\def\lambdaPCB{\hat{\boldsymbol{\lambda}}_{\mbox{\tiny MCP}}^{[b]}}

\def\hatuij{\hat{u}_{ij}}
\def\hatuonej{\hat{u}_{1j}}
\def\hatuzeroj{\hat{u}_{0j}}
\def\gonej{\boldsymbol{g}_{1j}(\theta)}
\def\gzeroj{\boldsymbol{g}_{0j}(\theta)}
\def\gij{\boldsymbol{g}_{ij}(\theta)}
\def\gijs{\boldsymbol{g}^{[b]}_{ij}(\theta)}
\def\gijzero{\boldsymbol{g}_{ij}(\theta^0)}
\def\gijhat{\boldsymbol{g}_{ij}\left(\thetaPC\right)}
\def\gijhatstar{\boldsymbol{g}^{[b]}_{ij}(\thetaPC)}

\def\aij{\Tilde{a}_{ij}}
\def\aijs{\Tilde{a}^{[b]}_{ij}}
\def\sumzo{\sum_{i=0}^1}

\def\mtbar{\bar{\hat{m}}_1}
\def\mcbar{\bar{\hat{m}}_0}

\def\Sigone{\sigma}
\def\Gam{\boldsymbol{\Gamma}}
\def\Wone{\boldsymbol{W}}
\def\Pone{\boldsymbol{P}_1}
\def\Ome{\boldsymbol{\Omega}}

\def\bfalpha{\mbox{\boldmath{$\alpha$}}}
\def\balpha{\boldsymbol{\alpha}}
\def\hbalpha{\hat{\boldsymbol{\alpha}}}

\newcommand{\indep}{\perp \!\!\! \perp}

\begin{document}

\begin{frontmatter}
\title{Pseudo-Empirical Likelihood Methods for Causal Inference}
\runtitle{PEL Methods for Causal Inference}

\begin{aug}
\author{\fnms{Jingyue} \snm{Huang}
\ead[label=e1]{jingyue.huang@uwaterloo.ca}}
\address{Department of Statistics and Actuarial Science\\
University of Waterloo, Waterloo, Ontario N2L 3G1, Canada\\
\printead{e1}}

\author{\fnms{Changbao} \snm{Wu}\thanksref{t2}\ead[label=e3]{cbwu@uwaterloo.ca}}
\address{Department of Statistics and Actuarial Science\\
University of Waterloo, Waterloo, Ontario N2L 3G1, Canada\\
\printead{e3}}

and

\author{\fnms{Leilei} \snm{Zeng}\thanksref{t3}\ead[label=e2]{lzeng@uwaterloo.ca}}
\address{Department of Statistics and Actuarial Science\\
University of Waterloo, Waterloo, Ontario N2L 3G1, Canada\\
\printead{e2}}

\thankstext{t2}{Supported in part by the Natural Sciences and Engineering Research Council of Canada and the Canadian Statistical Sciences Institute.}
\thankstext{t3}{Supported in part by the Natural Sciences and Engineering Research Council of Canada.}

\runauthor{J. Huang et al.}

\affiliation{University of Waterloo}

\end{aug}

\begin{abstract}
Causal inference problems have remained an important research topic over the past several decades due to their general applicability in assessing a treatment effect in many different real-world settings. 
In this paper, we propose two inferential procedures on the average treatment effect (ATE) through a two-sample pseudo-empirical likelihood (PEL) approach. The first procedure uses the estimated propensity scores for the formulation of the PEL function, and the resulting maximum PEL estimator of the ATE is equivalent to the inverse probability weighted estimator discussed in the literature. Our focus in this scenario is on the PEL ratio statistic and establishing its theoretical properties. The second procedure incorporates outcome regression models for PEL inference through model-calibration constraints, and the resulting maximum PEL estimator of the ATE is doubly robust. Our main theoretical result in this case is the establishment of the asymptotic distribution of the PEL ratio statistic. We also propose a bootstrap method for constructing PEL ratio confidence intervals for the ATE to bypass the scaling constant which is involved in the asymptotic distribution of the PEL ratio statistic but is very difficult to calculate. 
Finite sample performances of our proposed methods with comparisons to existing ones are investigated through simulation studies.

\end{abstract}

\begin{keyword}[class=MSC]
\kwd[Primary ]{62G05, 62G10}
\kwd[; Secondary ]{62G20}
\end{keyword}

\begin{keyword}
\kwd{Average treatment effect}
\kwd{propensity scores}
\kwd{pseudo-empirical likelihood ratio statistic}
\kwd{model calibration}
\kwd{double robustness}
\kwd{bootstrap method}
\end{keyword}
\tableofcontents
\end{frontmatter}

\section{Introduction}

It is straightforward to establish causal results when treatment assignments are completely randomized. It is known to the research community, however, that causal inferences with observational studies are a challenging task due to the non-randomized treatment assignments that are influenced by the covariates associated with outcomes. 
In such cases, the difference in outcomes between the treatment and the control groups is due to not only the different treatment exposures but also the different characteristics of units in the two groups as reflected by the covariates' imbalances. 
These covariates, often referred to as ``confounders", make the naive mean causal effect estimators invalid.

The propensity score, defined as the probability of being in the treatment group given the covariates, plays an essential role in causal inference for balancing covariates and valid statistical procedures, as highlighted in the seminal paper of \cite{Rosenbaum1983}.  
Many statistical methods are proposed based on propensity scores to obtain consistent estimators for the average treatment effect (ATE) under the so-called strongly ignorable assumption, such as matching \citep{Rosenbaum1985, Abadie2006}, post-stratification \citep{Rosenbaum1984, Rosenbaum1987}, and weighting \citep{Robins2000, Hirano2003}. 


A practical issue of the aforementioned methods is that the propensity scores are usually unknown and must be estimated. 
Misspecification of the propensity score model leads to invalid inverse probability weighted (IPW) estimators. 
One way to mitigate this issue is to construct estimators that combine inverse probability weighting with outcome regression modelling, known as the augmented inverse probability weighted (AIPW) estimators \citep{Robins1994}. 
The AIPW estimators are doubly robust in the sense that the consistency of such estimators only requires the correct specification of one of the two sets of models: the set of propensity score model and the set of outcome regression models \citep{Scharfstein1999}. 
The asymptotic variances of the AIPW estimators can be obtained based on their influence functions \citep{tsiatis2006semi}.  


Propensity score-based weighting methods can balance covariates in a large sample but may fail to do so in finite samples. 
Additionally, applied researchers often engage in a cyclical process of propensity score modelling and covariate balance checking until they achieve satisfactory results, which has been criticized as the ``propensity score tautology" \citep{Imai2008}.
To address these issues, \cite{Hainmueller2012} introduced entropy balancing to estimate the average treatment effect of the treated.
Entropy balancing exactly balances the sample moments of the covariates between the treatment and the control groups by maximizing the entropy of the weights subject to some calibration constraints that ensure the equivalence of the moments from the two groups.
\cite{Zhao2017} demonstrated that entropy balancing is doubly robust with respect to linear outcome regression and logistic propensity score regression, though there is no modelling in its original form. However, this approach may require a considerable number of constraints.

In this paper, we propose two procedures for the estimation and inference of the ATE through a two-sample pseudo-empirical likelihood (PEL) approach.
Two point estimators are constructed. 
One corresponds to the IPW estimator through the explicit use of estimated propensity scores in forming the PEL function, and the other achieves double robustness through the inclusion of additional model-calibration constraints based on the outcome regression models.
Moreover, for each procedure, we establish the asymptotic properties of the PEL ratio statistic, enabling the construction of confidence intervals and tests of hypotheses.
The methods we developed have attractive features shared by general empirical likelihood based approaches, which include
(1) range-respecting and transformation-invariant properties of the PEL ratio confidence intervals; and
(2) problem formulations through a constrained optimization procedure, which enables incorporations of suitable auxiliary information through additional constraints for more efficient or robust estimators \citep{Hall1990, owen2001}. 

The remainder of this paper is structured as follows.
Section \ref{section2} introduces fundamental concepts of causal inference, alongside commonly used causal inference methods.
In Section \ref{section3}, we present the formulation of the two-sample PEL approach to causal inference and develop the point estimator and the associated PEL ratio confidence intervals for the ATE. In Section \ref{section4}, we examine doubly robust inference under the proposed PEL framework through the inclusion of the model-calibration constraints. Results from simulation studies on finite sample performances of proposed methods with comparisons to existing ones are reported in Section \ref{section5}. Some additional remarks are given in Section \ref{section6}. Regularity conditions, proofs and technical details are presented in Section \ref{Sec.appendix}.

\section{Causal Inference}
\label{section2}

\subsection{Basic setting and propensity scores}

We follow the commonly used setup of causal inference with the potential outcome framework. 
Let $T$ be a binary variable denoting the treatment assignment, with $T=1$ indicating treatment and $T=0$ for control. 
The potential outcome variable under the treatment or the control is represented by $Y_1$ or $Y_0$, respectively.
The parameter of interest is the average treatment effect (ATE), $\theta=\mu_1-\mu_0$, where $\mu_1= \operatorname{E}(Y_1)$ and $\mu_0=\operatorname{E}(Y_0)$ are the expectations of the potential outcome variables. 
As each individual can only be assigned to one of the two groups, only one of the potential outcomes can be observed for each individual, and the observed outcome is denoted as  $Y=TY_1+(1-T)Y_0$. 

Let $T_j$, $Y_{1j}$ and $Y_{0j}$ be respectively the values of $T$, $Y_1$ and $Y_0$ associated with subject $j$. Let $Y_j=T_jY_{1j}+(1-T_j)Y_{0j}$. 
Consider a random sample $\bs$ of size $n$ from an infinite target population. 
The sample data are represented by $\{(\bx, Y_j, \Ri), j\in \bs \}$, where $\bx$ denotes the value of the vector of auxiliary variables for subject $j$, $j=1, \dots, n$. 
We define $\bst=\{j \mid \Ri=1 \text{ and } j\in \bs\}$ and $\bsc=\{j \mid \Ri=0\text{ and }j\in\bs\}$ as the sub-samples of subjects receiving treatment and control, respectively.
The available sample data can be partitioned into two subsets: $\{(\bx, \Yti, \Ri=1), j\in \bst \}$ and $\{(\bx, \Yci, \Ri=0), j\in \bsc \}$.
Denote the sizes of $\bst$ and $\bsc$ respectively by $n_1$ and $n_0$. We have $n=n_1+n_0$.

Balancing the distributions of confounders, covariates that are simultaneously associated with the treatment and potential response variables, in the two treatment groups is essential for investigating causal effects in observational studies. 
The propensity scores are a widely used tool for this purpose. 
The propensity score (PS) is defined as the conditional probability of receiving the treatment given the covariates, denoted as $P(T=1\mid \boldsymbol{x})$. 
The properties and the essential role of propensity scores in causal inference are elaborated in the seminal paper by \cite{Rosenbaum1983}. However, it is important to note that certain critical assumptions must be met in order to obtain consistent estimators of causal effects through propensity score adjustments. 
Two key assumptions are given below. 
\begin{description}
\item[A1.] {\em Strongly Ignorable Treatment Assignment (SITA).} The treatment indicator ($T$) and the potential response variables ($Y_1$, $Y_0$) are independent given the set of covariates ($\boldsymbol{x}$).
\item[A2.] {\em Positivity Assumption.} The propensity scores are strictly in the range of $(0,1)$, i.e., $0<P(T=1\mid\boldsymbol{x})<1$ for all possible values of $\boldsymbol{x}$.
\end{description}
The SITA assumption {\bf A1} implies that there are no unmeasured confounders and treatment assignments depend only on the observed covariates. The positivity assumption {\bf A2} guarantees that there is no restriction preventing individuals from being assigned to either the treatment or the control group.

Propensity scores are typically unknown and require to be estimated in practice.
Under a parametric model, we have the propensity score $\tau_j = P(T_j=1\mid \boldsymbol{x}_j)=\tau(\boldsymbol{x}_j; \bfalpha)$, where $\tau(\cdot;\cdot)$ has a known form and $\bfalpha$ is the vector of unknown parameters. 
The maximum likelihood estimator $\hat{\boldsymbol{\alpha}}$ of $\boldsymbol{\alpha}$ can be obtained by maximizing the likelihood function $L(\bfalpha)=\prod_{j=1}^n\{\tau(\boldsymbol{x}_j; \bfalpha)\}^{T_j}\{1-\tau(\boldsymbol{x}_j; \bfalpha)\}^{1-T_j}$ using the observed dataset $\{(\bx, \Ri), j\in \bs \}$, which yields estimated propensity scores $\htau=\tau(\bx;\hbalpha)$, $j=1, \dots, n$.
In practice, logistic regression models are commonly employed for the propensity scores, leading to the following closed-form expressions for the estimated propensity scores, 
\[
    \htau=\tau(\bx;\hbalpha)=\text{expit}\left(\bz^\top\hbalpha\right)=\frac{\exp\left(\bz^\top\hbalpha\right)}{1+\exp\left(\bz^\top\hbalpha\right)}\, ,
    \label{tau}
\]
where $\bz=(1, \bx^\top)^\top$. The discussions in the rest of this paper assume a logistic regression model for propensity scores.

\subsection{Inverse probability weighted estimators}
\label{section-ipw}

The method of inverse probability weighting was originally introduced by \cite{horvitz1952thompson} for estimating a finite population total using a probability survey sample, where the weighting is done through the inverses of the known sample inclusion probabilities. 
The resulting weighted estimator is called the Horvitz-Thompson (HT) estimator, which is one of the backbones for design-based inference in survey sampling. The concept was successfully adapted to causal inference and missing data analysis based on estimated propensity scores. 

The inverse probability weighted (IPW) estimators, also referred to as the propensity score-adjusted estimators, play a crucial role in causal inference. The IPW estimators of $\mu_1$ and $\mu_0$ are given by
\[
\scalemath{0.92}{
    \mutIPW = \frac{1}{n} \sbsj \frac{\Ri Y_j }{\htau}=\frac{1}{n} \sbstj\frac{\Yti}{\htau}} \;\;{\rm and} \;\;
    \scalemath{0.92}{
\mucIPW = \frac{1}{n} \sbsj \frac{\left(1-\Ri\right) Y_j}{1-\htau}=\frac{1}{n} \sbscj\frac{\Yci}{1-\htau}}\, .
\]
The corresponding IPW estimator of the ATE is defined as $\thetaIPW=\mutIPW-\mucIPW$.

Note that $\operatorname{E}(\sum_{l\in\bst} {\tau}_l^{-1})={\operatorname{E}(\sum_{l\in \bs} T_l {\tau}_l^{-1})}=n$. Replacing $n$ by its consistent estimator $\sum_{l\in\bst} \hat{\tau}_l^{-1}$ in $\mutIPW$ yields the H\'{a}jek-type \citep{hajek1971estimator} IPW estimator for $\mu_1$, 
\begin{equation}
    \mutH = \sbstj \frac{{\htau}^{-1}}{\sum_{l\in\bst} \hat{\tau}_l^{-1}}  {\Yti}
    =\sbstj \Tilde{a}_{1j} {\Yti}\, ,
    \label{mu1IPW2}
\end{equation}
where ${\tilde{a}_{1j} = \hat{\tau}_j^{{-1}}/\sum_{l\in \bst}\hat{\tau}_l^{{-1}}}$ and $\sum_{j\in \bst}\tilde{a}_{1j} = 1$. 
Similarly, the H\'{a}jek-type IPW estimator for $\mu_0$ takes the form 
\begin{equation}
 \mucH =  \sbscj \frac{{(1-\htau)}^{-1}}{\sum_{l\in\bsc} (1-\hat{\tau}_l)^{-1}} {\Yci}
=\sbscj \Tilde{a}_{0j} \Yci\, ,   
\label{mu0IPW2}
\end{equation}
where $\tilde{a}_{0j} = (1-\hat{\tau}_j)^{{-1}}/\sum_{l\in \bsc}(1-\hat{\tau}_l)^{{-1}}$.
The resulting estimator for the ATE is then given by $\thetaH=\mutH-\mucH$.

The IPW estimators $\thetaIPW$ and $\thetaH$ of the ATE are consistent under the assumed propensity score model and certain regularity conditions. 
This can be shown by recognizing that they belong to the so-called $m$-estimators and following the technical arguments presented in Section 3.2 of \cite{tsiatis2006semi}. Furthermore, the variance of the IPW estimators can be estimated by the sandwich variance estimator. Details are presented in Section \ref{sandforIPW}. The H\'{a}jek-type IPW estimator $\thetaH$ is typically more efficient than the IPW estimator $\thetaIPW$ for finite samples \citep{sarndal1992survey}.  

\subsection{Augmented IPW estimators}

The IPW estimators can be biased when the propensity score model is misspecified. The biases can sometimes be mitigated through the incorporation of an outcome regression model. 
\cite{Robins1994} proposed a class of augmented IPW estimators under the two-model framework, which usually exhibits greater efficiency than the IPW estimators when both models are correct. 
\cite{Scharfstein1999} later found that the consistency of these estimators only requires 
one of the two models to be correctly specified. 
Hence, such estimators are also termed as doubly robust. 

Under the potential outcome framework, consider two outcome regression models: $\operatorname{E}(Y_{1}\mid \boldsymbol{x})=m_1(\boldsymbol{x};\boldsymbol{\beta}_1)$ and $\operatorname{E}(Y_{0}\mid \boldsymbol{x})=m_0(\boldsymbol{x};\boldsymbol{\beta}_0)$. 
Given the SITA assumption {\bf A1}, i.e., $T \indep \{Y_1, Y_0\}\mid \boldsymbol{x}$, we have 
\[
  \operatorname{E}(Y_1\mid \boldsymbol{x})=\operatorname{E}(Y_1\mid T=1, \boldsymbol{x})  \;\;\;{\rm and} \;\;\; \operatorname{E}(Y_0\mid \boldsymbol{x})=\operatorname{E}(Y_0\mid T=0, \boldsymbol{x}) \, .
\]
It follows that the model parameters $\boldsymbol{\beta}_1$ and $\boldsymbol{\beta}_0$ can be estimated by fitting the corresponding model using observed datasets $\{(\boldsymbol{x}_j,Y_{1j}),j\in \bst\}$ and $\{(\boldsymbol{x}_j,Y_{0j}),$\\$j\in \bsc\}$, respectively.
Let $\hat{m}_{1j}=m_1(\boldsymbol{x}_j; \hat{\boldsymbol{\beta}}_1)$ and $\hat{m}_{0j}=m_0(\boldsymbol{x}_j; \hat{\boldsymbol{\beta}}_0)$, for all $j\in\bs$, where $\hat{\boldsymbol{\beta}}_1$ and $\hat{\boldsymbol{\beta}}_0$ are the corresponding estimators of $\boldsymbol{\beta}_1$ and $\boldsymbol{\beta}_0$.

Following \cite{Robins1994}, the augmented IPW estimators for $\mu_1$ and $\mu_0$ are constructed as 
\begin{eqnarray*}
\mutDRO &=& \frac{1}{n} \sum_{j \in \bst} \frac{ Y_{1j}- \hat{m}_{1j}}{\hat{\tau}_{j}} + \frac{1}{n} \sum_{j=1}^{n}\hat{m}_{1j} \,, \\
\mucDRO &=& \frac{1}{n} \sum_{j \in \bsc} \frac{ Y_{0j}- \hat{m}_{0j}}{1-\hat{\tau}_{j}} + \frac{1}{n} \sum_{j=1}^{n}\hat{m}_{0j} \,,
\end{eqnarray*}
and the augmented IPW estimator for the ATE is given by 
$\thetaDRO = \mutDRO-\mucDRO$.

Each of the two augmented IPW estimators, $\mutDRO$ and  $\mucDRO$, contains an IPW estimator based on the ``residuals'' $Y_{1j}- \hat{m}_{1j}$ or $Y_{0j}- \hat{m}_{0j}$, and each IPW estimator can be replaced by a H\'{a}jek-type IPW estimator, leading to 
\[
\mutDRT = \sbstj \Tilde{a}_{1j}\left(Y_{1j}-\hat{m}_{1j}\right)+\mtbar \;\; {\rm and} \;\;
\mucDRT = \sbscj\Tilde{a}_{0j}\left(Y_{0j}-\hat{m}_{0j}\right)+\mcbar\,,
\]
where $\mtbar=\sum_{j=1}^n\hat{m}_{1j}/n$, $\mcbar=\sum_{j=1}^n\hat{m}_{0j}/n$, $\Tilde{a}_{1j}$ and  $\Tilde{a}_{0j}$ are defined in (\ref{mu1IPW2}) and (\ref{mu0IPW2}). The corresponding estimator for the ATE is given by $ \thetaDRT = \mutDRT - \mucDRT$.

\begin{remark}
    The estimators $\thetaDRO$ and $\thetaDRT$ are doubly robust in the sense that they are consistent estimators of $\theta$ if one of the two sets of models, the propensity score model $\tau(\boldsymbol{x};\balpha)$ or the set of outcome regression models $m_1(\boldsymbol{x};\boldsymbol{\beta}_1)$ and $m_0(\boldsymbol{x};\boldsymbol{\beta}_0)$, is correctly specified. This is clearly an attractive property that offers some protection against misspecification of one set of models. 
\end{remark}

Doubly robust variance estimation based on a doubly robust point estimator is an active research topic pursued by several authors. 
In the context of missing data, \cite{Cao2009} briefly mentioned that the variance of some augmented IPW estimators for the population mean can be estimated by the usual empirical sandwich technique \citep{Stefanski2002} if the point estimators are obtained by solving a set of $m$-estimating equations jointly.  The resulting variance estimator is robust against misspecifications of one or both of the propensity score and outcome regression models.
In the current setting, the sandwich variance estimators for the two augmented IPW estimators, $\thetaDRO$ and $\thetaDRT$, are consistent for the true variance even if one or both of the two sets of models are incorrectly specified. 
The sandwich variance estimators for the augmented IPW estimators are constructed in the same manner as those for the IPW estimators, where the additional estimating equations for the outcome regression models are included.
We formally state this result in the following proposition. Proof of the result is presented in Section \ref{proofAIPW}, where we also present the detailed formulation of the sandwich variance estimator for the augmented IPW estimator.

\begin{corollary}
Under certain regularity conditions, the variance of the augmented IPW estimator $\thetaDRO$ or $\thetaDRT$ can be estimated by the sandwich variance estimator regardless of the correctness of the propensity score model or the outcome regression models.
\label{corollary1}
\end{corollary}

The required regularity conditions for Proposition \ref{corollary1} as well as other major results presented in the paper are given in Section \ref{RC}.
When both sets of models are misspecified, the point estimators become invalid, and the corresponding variance estimators are not practically useful.
We mainly focus on the doubly robust property of the sandwich variance estimators for the augmented IPW estimators.
We also demonstrate that a version of bootstrap variance estimators is doubly robust and can provide valid standard errors when either the outcome regression models or the propensity score model is misspecified, through simulation studies in Section \ref{section5}.
Let $\thetaDR$ denote $\thetaDRO$ or $\thetaDRT$. The algorithm for obtaining the bootstrap variance estimator is described as follows:
\begin{description}

\item[Step 1.] Select a bootstrap sample $\mathcal{S}^{[b]}$ of size $n$ from the original sample $\mathcal{S}$ using simple random sampling with replacement.
    
\item[Step 2.] Calculate the bootstrap version of the doubly robust estimator, denoted as $\thetaDR^{[b]}$, based on the bootstrap sample $\mathcal{S}^{[b]}$.

\item[Step 3.] Repeat Steps 1 and 2 a large number $B$ times, independently, to obtain $\thetaDR^{[b]}$, $b=1,2,\cdots,B$. 
    
\item[Step 4.] Compute the bootstrap variance estimator of $\thetaDR$ as
\[
\operatorname{var}\left(\thetaDR\right)=\frac{1}{B}\sum_{b=1}^B\left(\thetaDR^{[b]}-\bar{\hat{\theta}}_{\mbox{\tiny DR}}\right)^2\, ,
\] 
where $\bar{\hat{\theta}}_{\mbox{\tiny DR}}=B^{-1}\sum_{b=1}^B\thetaDR^{[b]}$.

\end{description}

\section{The PEL Approach to Causal Inference }
\label{section3}


\subsection{Pseudo-empirical likelihood}

Empirical likelihood methods, first proposed by \cite{Owen1988}, are a powerful nonparametric tool for statistical inference under the likelihood principle analogous to parametric likelihood methods. 
Maximum empirical likelihood estimators are obtained through constrained maximization of the empirical likelihood function, and confidence intervals and hypothesis tests can be constructed through the empirical likelihood ratio function. 
\cite{Owen1988} first established the empirical likelihood theorem, a nonparametric version of the Wilks' Theorem \citep{Wilks1938}, 
and generalized the result to multivariate functionals \citep{Owen1990} and to linear regression models \citep{Owen1991}. 
\cite{Chen1993} applied the empirical likelihood methods to finite population parameters under simple random sampling for more efficient estimation through the inclusion of additional constraints using auxiliary information. 
\cite{Qin1994} provided a general framework by combining empirical likelihood methods with unbiased estimating equations. 

\cite{Chen1999} proposed the pseudo-empirical likelihood (PEL) approach to complex survey data, with the survey weights incorporated into the formulation of the empirical likelihood, to provide design-consistent point estimators for finite population parameters.  
\cite{Wu2006} developed an alternative form of the PEL function for complex survey data to construct PEL ratio confidence intervals that retain all the attractive features of the empirical likelihood methods. The PEL function of \cite{Wu2006} serves as the starting point of our proposed methods for causal inference. In this section, we focus on PEL methods for the ATE under an assumed propensity score model. A doubly robust PEL approach to causal inference is discussed in Section \ref{section4}. 

\subsection{PEL estimation of the ATE}
\label{PELSection}

The pseudo-empirical likelihood function presented in \cite{Wu2006} in the context of survey sampling can be adapted for causal inference problems. We first replace the sampling weights by suitable replacements using the estimated propensity scores for subjects in $\bst$ and $\bsc$, and then construct the PEL function using a two-sample empirical likelihood formulation as in \cite{Wu2012}. Our proposed joint PEL function for the two samples, $\bst$ and $\bsc$, is given by 
\begin{equation}
\ell_{\mbox{\tiny PEL}}\left(\boldsymbol{p}_{1}, \boldsymbol{p}_{0}\right)= n\left\{w_{1} \sum_{j\in \bst} \tilde{a}_{1 j} \log \left(p_{1 j}\right) +w_{0} \sum_{j\in \bsc} \tilde{a}_{0 j} \log \left(p_{0 j}\right)\right\} \; ,
\label{jointPEL}
\end{equation}
where $\mbox{\boldmath{$p$}}_i = (p_{i1},\cdots,p_{in_i})^\top$ is a set of discrete probability measure imposed over $\bs_i$ for $i=0,1$, $w_1=w_0=1/2$, and the normalized weights $\tilde{a}_{1j}$ and $\tilde{a}_{0j}$ are defined in Section \ref{section-ipw} using the estimated propensity scores. The use of $w_1=w_0=1/2$ in the PEL function $\ell_{\mbox{\tiny PEL}}\left(\boldsymbol{p}_{1}, \boldsymbol{p}_{0}\right)$ is to mimic the PEL function of \cite{Wu2006} for stratified sampling, which leads to a simple computational procedure. 
Maximizing $\ell_{\mbox{\tiny PEL}}(\boldsymbol{p}_{1}, \boldsymbol{p}_{0})$ in (\ref{jointPEL}) subject to the normalization constraints 
\begin{equation}
    \sum_{j\in\bst}p_{1j}=1  \;\;\;{\rm and} \;\;\; \sum_{j\in\bsc}p_{0j}=1
    \tag{C1}
    \label{jnorm}    
\end{equation}
yields $\hat{\mbox{\boldmath{$p$}}}_i=(\hat{p}_{i1},\dots, \hat{p}_{in_i})^\top$, where $\hat{p}_{ij} = \tilde{a}_{ij}$ for $j\in \bs_i$, $i=1,0$.  
The maximum PEL estimator of $\mu_i$ is computed as 
\[
    \hat{\mu}_{i\mbox{\tiny PEL}} = \sum_{j\in \bs_i}\hat{p}_{ij} Y_{ij}={\sum_{j\in \bs_i} \tilde{a}_{i j}  Y_{ij}}
\]
for $i=1$ and $0$, which is identical to the H\'{a}jek-type IPW estimators of $\mu_i$ given in (\ref{mu1IPW2}) and (\ref{mu0IPW2}).  
The maximum PEL estimator of the ATE is given by $\thetaP=\hat{\mu}_{1\mbox{\tiny PEL}}-\hat{\mu}_{0\mbox{\tiny PEL}}$, which is also the same as the H\'{a}jek-type IPW estimator $\thetaH$.

\subsection{PEL ratio confidence intervals}
\label{PEL-CI}

One major advantage of the PEL approach is that we can construct the PEL ratio statistic for the parameter of interest, the ATE, as well as other parameters.
Studying the asymptotic properties of the PEL ratio statistic enables the construction of confidence intervals and hypothesis tests.
The ``global'' maximum of the PEL function $\ell_{\mbox{\tiny PEL}}({\mbox{\boldmath{$p$}}}_1, {\mbox{\boldmath{$p$}}}_0)$ under the normalization constraints (\ref{jnorm}) is achieved at $\hat{\mbox{\boldmath{$p$}}}_i=(\hat{p}_{i1},\dots, \hat{p}_{in_i})^\top$, where $\hat{p}_{ij} = \tilde{a}_{ij}$ for $j\in \bs_i$, $i=1,0$,
as shown in Section \ref{PELSection}. 
Let $\scalemath{0.97}{{\hat{\mbox{\boldmath{$p$}}}_1(\theta) = (\hat{p}_{11}(\theta),\cdots,\hat{p}_{1n_1}(\theta))^\top}}$ and $\scalemath{0.97}{{\hat{\mbox{\boldmath{$p$}}}_0(\theta) =} {(\hat{p}_{01}(\theta),\cdots,\hat{p}_{0n_0}(\theta))^\top}}$ be the ``restricted'' maximizer of $\ell_{\mbox{\tiny PEL}}({\mbox{\boldmath{$p$}}}_1, {\mbox{\boldmath{$p$}}}_0)$ under the normalization constraints (\ref{jnorm}) and the following constraint (\ref{para}) induced by the parameter of interest, $\theta=\mu_1-\mu_0$, 
\begin{equation}
    \sum_{j\in \bst}p_{1j}Y_{1j}-\sum_{j\in \bsc}p_{0j}Y_{0j}=\theta
    \tag{C2}
    \label{para}
\end{equation}
for a given $\theta$. The profile PEL function for $\theta$ is then given by 
\begin{equation}
\scalemath{0.97}{
\ell_{\mbox{\tiny PEL}}(\hat{\mbox{\boldmath{$p$}}}_1(\theta), \hat{\mbox{\boldmath{$p$}}}_0(\theta)) = n\left\{w_1\sum_{j\in \bst}\tilde{a}_{1j}\log(\hat{p}_{1j}(\theta))+w_0\sum_{j\in \bsc}\tilde{a}_{0j}\log(\hat{p}_{0j}(\theta))\right\}}\, .   
\label{PELtheta}
\end{equation}
The maximum PEL estimator of the ATE $\theta$ can be alternatively defined as the maximizer of the profile PEL function $\ell_{\mbox{\tiny PEL}}(\hat{\mbox{\boldmath{$p$}}}_1(\theta), \hat{\mbox{\boldmath{$p$}}}_0(\theta))$ with respect to $\theta$. This alternative definition is equivalent to $\thetaP=\hat{\mu}_{1\mbox{\tiny PEL}}-\hat{\mu}_{0\mbox{\tiny PEL}}$ presented in Section \ref{PELSection}. 

\begin{lemma}
The maximum PEL estimator $\thetaP =\hat{\mu}_{1\mbox{\tiny PEL}}-\hat{\mu}_{0\mbox{\tiny PEL}}$ maximizes the profile PEL function $\ell_{\mbox{\tiny PEL}}(\hat{\mbox{\boldmath{$p$}}}_1(\theta), \hat{\mbox{\boldmath{$p$}}}_0(\theta))$ given in (\ref{PELtheta}) with respect to $\theta$. 
\end{lemma}

We now present our first major result on the PEL ratio statistic for the ATE,  $\theta$. For clarity of presentation, we let $\mu_1^0$ and $\mu_0^0$ denote the true values of the population means $\mu_1$ and $\mu_0$, and $\theta^0=\mu_1^0-\mu_0^0$ be the true value of the ATE. 
The PEL ratio function for $\theta$ is defined as 
\begin{equation}
    r_{\mbox{\tiny PEL}}(\theta)=\ell_{\mbox{\tiny PEL}}(\hat{\mbox{\boldmath{$p$}}}_1(\theta), \hat{\mbox{\boldmath{$p$}}}_0(\theta))-\ell_{\mbox{\tiny PEL}}(\hat{\mbox{\boldmath{$p$}}}_1, \hat{\mbox{\boldmath{$p$}}}_0)\, ,
    \label{PELratio}
\end{equation}  
where $(\hat{\mbox{\boldmath{$p$}}}_1, \hat{\mbox{\boldmath{$p$}}}_0)$ and $(\hat{\mbox{\boldmath{$p$}}}_1(\theta), \hat{\mbox{\boldmath{$p$}}}_0(\theta))$ are the ``global'' maximizer and the ``restricted'' maximizer of the PEL function $\ell_{\mbox{\tiny PEL}}\left(\boldsymbol{p}_{1}, \boldsymbol{p}_{0}\right)$ for a given $\theta$, respectively. 
Let $\chi^2_1$ denote a chi-squared distribution with one degree of freedom.
The following theorem states the limiting distribution of the PEL ratio statistic $r_{\mbox{\tiny PEL}}(\theta)$ at $\theta=\theta^0$. 

\begin{theorem}
Under certain regularity conditions and a correctly specified propensity score model, the scaled PEL ratio function $-2r_{\mbox{\tiny PEL}}(\theta)/\hat{c}$ at $\theta=\theta^0$ converges in distribution to a $\chi^2_1$ distributed random variable as $n\rightarrow\infty$, where the adjusting factor $\hat{c}$ is given in (\ref{scalecons}).
\label{PELWOMC}
\end{theorem}

The scaling constant can be theoretically defined at the population level as $c$. In applications, it suffices to use a consistent estimator $\hat{c}$ of $c$, which is given by
\begin{equation}
\scalemath{0.98}{
\hat{c}=n\left\{2\left[\sum_{j\in \bst}\tilde{a}_{1j}Y_{1j}^2+\sum_{j\in \bsc}\tilde{a}_{0j}Y_{0j}^2-\mutH^2-\mucH^2\right]\right\}^{-1}\operatorname{var}\left(\thetaH\right)}\, , 
\label{scalecons}
\end{equation}
where $\operatorname{var}(\thetaH)$ is the variance estimator of the H\`{a}jek-type IPW estimator $\thetaH$.
The explicit expression of $\operatorname{var}(\thetaH)$ is given in Section \ref{sandforIPW}. 
Proof of Theorem \ref{PELWOMC} is presented in Section \ref{PELWOMCProof}.

Using the result of Theorem \ref{PELWOMC}, we can construct a $100(1-\alpha)\%$-level PEL ratio confidence interval for $\theta$ in the form of  
\begin{equation}
    \{\theta \, | \, -2r_{\mbox{\tiny PEL}}(\theta)/\hat{c}\leq \chi^2_1(\alpha)\}\; ,
    \notag
\end{equation}
where $\chi^2_1(\alpha)$ is the $100(1-\alpha)$th quantile of the $\chi^2_1$ distribution for a given $\alpha\in(0,1)$. \cite{Wu2005} contains algorithmic details on how to find the PEL ratio confidence interval using a simple bi-section search method.

\section{The PEL Approach to Doubly Robust Estimation}
\label{section4}


The PEL approach proposed in Section \ref{section3} to causal inference can be further extended to achieve doubly robust estimation through the inclusion of model-calibration constraints. Calibration methods were first developed in survey sampling \citep{Deville1992}, where calibration constraints were imposed directly over individual auxiliary variables with known population controls. 
\cite{Wu2001} proposed a model-calibration technique based on a working linear or nonlinear model, where a single constraint is formed based on fitted values. The model-calibration estimators were shown to be optimal among a class of calibration estimators \citep{Wu2003}. In this section, we demonstrate that outcome regression models can be used to form model-calibration constraints to achieve doubly robust estimation for the ATE.

\subsection{Model-calibrated maximum PEL estimators}
\label{MPELWMCsection}

We consider two model-calibration constraints using the two outcome regression models. Note that 
$\hat{m}_{1j}=m_1(\boldsymbol{x}_j; \hat{\boldsymbol{\beta}}_1)$, $\hat{m}_{0j}=m_0(\boldsymbol{x}_j; \hat{\boldsymbol{\beta}}_0)$, 
$\mtbar=\sum_{j=1}^n\hat{m}_{1j}/n$, and $\mcbar=\sum_{j=1}^n\hat{m}_{0j}/n$. The two constraints on the empirical probabilities 
$\mbox{\boldmath{$p$}}_i = (p_{i1},\cdots,p_{in_i})^\top$ are formed as  
\begin{equation}
    \sum_{j \in \bst} p_{1j}\hat{m}_{1j}=\mtbar     \;\;\;{\rm and} \;\;\;    \sum_{j \in \bsc} p_{0j}\hat{m}_{0j}=\mcbar \,.
    \label{jcal}
    \tag{C3}
\end{equation}
The model-calibrated maximum PEL estimator of $\mu_i$ is computed as $\hat{\mu}_{i\mbox{\tiny MCP}}=\sum_{j\in \bs_i}\hat{p}_{ij} Y_{ij}$, where 
$\hat{\mbox{\boldmath{$p$}}}_i = (\hat{p}_{i1},\cdots,\hat{p}_{in_i})^\top$, $i=0,1$ maximizes the PEL function 
$\ell_{\mbox{\tiny PEL}}\left(\boldsymbol{p}_{1}, \boldsymbol{p}_{0}\right)$ 
subject to the normalization constraints (\ref{jnorm}) and the model-calibration constraints (\ref{jcal}). 
It can be shown by using the Lagrange multiplier method that the constrained maximizers are given by 
$\hat{p}_{ij}=\tilde{a}_{ij}/(1+\hat{\lambda}_i\hat{u}_{ij})$, where $\hat{\lambda}_i$ is the solution to
\[
\sum_{j\in \bs_i} \frac{\tilde{a}_{ij}\hat{u}_{ij}}{1+\lambda_i \hat{u}_{ij}}=0\,, \;\;\; i=1,0\,,
\]
and $\hat{u}_{ij}=\hat{m}_{ij}-\Bar{\hat{m}}_i$. The asymptotic properties of $\hat{\mu}_{1\mbox{\tiny MCP}}$ are summarized in the theorem below. Parallel results can be stated regarding $\hat{\mu}_{0\mbox{\tiny MCP}}$.

\begin{theorem}
Under suitable regularity conditions, the model-calibrated maximum PEL estimator $\hat{\mu}_{1\mbox{\tiny MCP}}$ of $\mu_1$ is doubly robust with respect to the propensity score model and the outcome regression model. Moreover, if the propensity score model is correctly specified,  the estimator $\hat{\mu}_{1\mbox{\tiny MCP}}$ admits the asymptotic expansion 
\begin{equation}
    \hat{\mu}_{1\mbox{\tiny MCP}} =\sum_{j \in \bst} \tilde{a}_{1j} Y_{1j}+\hat{B}\left\{\mtbar-\sum_{j \in \bst} \tilde{a}_{1j} \hat{m}_{1j}\right\}+o_{p}\left(n^{-1 / 2}\right)\; ,  
    \label{mutPC}
\end{equation}
where 
\be
\hat{B}=\left\{\sum_{j \in \bst} \tilde{a}_{1j}\hat{u}_{1j} Y_{1j}\right\}  \left\{\sum_{j \in \bst} \tilde{a}_{1j}\hat{u}_{1j}^{2}\right\}^{-1}\, .
\label{mutPC-B}
\ee
\label{mutPCch2}
\end{theorem}

The inclusion of the model-calibration constraints leads to a doubly robust estimator for the ATE as $\thetaPC=\hat{\mu}_{1\mbox{\tiny MCP}}-\hat{\mu}_{0\mbox{\tiny MCP}}$. This estimator is more efficient than the IPW estimator when both sets of models are correctly specified. 
Specifically, if the outcome regression model is correctly specified, the $\hat{B}$ given in (\ref{mutPC-B}) converges to $1$, and the newly constructed estimator $\hat{\mu}_{1\mbox{\tiny MCP}}$ is asymptotically equivalent to the augmented IPW estimator $\mutDRT$ and is particularly efficient. The $\hat{B}$ does not converge to $1$ otherwise.
The expression (\ref{mutPC}) also implies that $\hat{\mu}_{1\mbox{\tiny MCP}}=\mutH+O_p(n^{-1/2})$, which further implies $\hat{\mu}_{1\mbox{\tiny MCP}}-\mu_1^0=O_p(n^{-1/2})$ when the propensity score model is correctly specified. 
Proof of Theorem \ref{mutPCch2} is given in Section \ref{mutPCch2proof}.

An important feature of our proposed PEL approach to causal inference is the flexibility of including additional constraints under the constrained maximization framework. Useful information can be incorporated through suitable constraints, which often lead to more robust and more efficient estimators \citep{Chan2014}. 
This is related to a topic of recent research interests, the so-called multiply robust estimators; 
see {\cite{Han2013}}, \cite{Han2014a}, \cite{Han2014b}, \cite{Han2016}, \cite{Chen2017}, and \cite{Duan2017}, among others, for further discussion.

\subsection{Model-calibrated PEL ratio confidence intervals}
\label{PELRBCI}

We now establish the limiting distribution of the PEL ratio statistic with the inclusion of model-calibration constraints. Let $\hat{\mbox{\boldmath{$p$}}}_1(\theta) = (\hat{p}_{11}(\theta),\cdots,\hat{p}_{1n_1}(\theta))^\top$ and 
$\hat{\mbox{\boldmath{$p$}}}_0(\theta) = (\hat{p}_{01}(\theta),\cdots,\hat{p}_{0n_0}(\theta))^\top$ be the maximizer of the PEL function $\ell_{\mbox{\tiny PEL}}\left(\boldsymbol{p}_{1}, \boldsymbol{p}_{0}\right)$ given in ({\ref{jointPEL}}) under the normalization constraints (\ref{jnorm}), the model-calibration constraints (\ref{jcal}), and the parameter constraint (\ref{para}) for a fixed $\theta$. 
Let $\ell_{\mbox{\tiny PEL}}(\theta) = \ell_{\mbox{\tiny PEL}}\left(\hat{\boldsymbol{p}}_{1}(\theta), \hat{\boldsymbol{p}}_{0}(\theta)\right)$ be the profile PEL function of $\theta$, which is the ``restricted'' maximum of the PEL function under the constraints (\ref{jnorm}), (\ref{jcal}) and (\ref{para}) for the given $\theta$. 

\begin{lemma}
The model-calibrated maximum PEL estimator $\thetaPC$ maximizes the profile PEL function $\ell_{\mbox{\tiny PEL}}(\theta)$ with respect to $\theta$. 
\end{lemma}

The profile PEL function $\ell_{\mbox{\tiny PEL}}(\theta)$ with a given $\theta$ can be computed by existing algorithms for empirical likelihood through reformulations of the constraints. 
Under the normalization constraints (\ref{jnorm}), the model calibration constraints (\ref{jcal}) can be rewritten as $\sum_{j\in\bs_i}p_{ij}\hatuij=0$ for $i=1,0$,
and the parameter constraint (\ref{para}) is equivalent to $\sbstj p_{1j}r_{1j}-\sbscj p_{0j}r_{0j}=0$, where $r_{1j}=Y_{1j}-\theta/2$ and $r_{0j}=Y_{0j}+\theta/2$.
Note that $w_1=w_0=1/2$. The constraints (\ref{jnorm})-(\ref{jcal}) altogether can be thus reformulated as 
\begin{equation}
\begin{array}{l}\sum_{i=0}^{1} w_{i} \sum_{j=1}^{n_{i}} p_{i j}=1 \\ \sum_{i=0}^{1} w_{i} \sum_{j=1}^{n_{i}} p_{i j} \boldsymbol{g}_{i j}(\theta)={\bf 0}\; ,
\end{array}
\tag{C4}
\label{reformulatedWith}
\end{equation}
where $\scalemath{0.93}{\gonej=(1-w_1, \hatuonej, \hatuonej, r_{1j}/w_1)^\top}$ and $\scalemath{0.93}{\gzeroj=(-w_1, \hatuzeroj, -\hatuzeroj, -r_{0j}/w_0)^\top}$. 

The major motivation behind this reformulation is to use computational procedures for stratified sampling as described in \cite{Wu2006} and \cite{Wu2005}. 
Maximizing (\ref{jointPEL}) subject to the constraints in (\ref{reformulatedWith}) for a fixed $\theta$ yields the solution \[\hat{p}_{ij}(\theta)=\frac{\aij}{1+\hat{\boldsymbol{\lambda}}^\top\gij}\; ,\] 
where the Lagrange multiplier $\hat{\boldsymbol{\lambda}}$ is obtained by solving
\[
\sumzo w_i \sbsji \frac{\aij \gij}{1+\boldsymbol{\lambda}^\top \gij} = {\bf 0}\; .
\]

Let $\Wone$ be the limit of $\sumzo w_i\sbsji\tilde{a}_{ij}\gijzero\gijzero^\top$, $\Gam=(0,0,0,-1)^\top$, and $\Sigone=(\Gam^\top\Wone^{-1}\Gam)^{-1}$. We denote the asymptotic variance-covariance matrix of $\scalemath{0.84}{\textstyle \sqrt{n}\sumzo}$ $w_i \sbsji \aij \gijzero$ by $\Ome$.
In what follows, we present the limiting distribution of the maximum PEL estimator $\thetaPC$ in Theorem \ref{normalPoint} and the limiting distribution of the PEL ratio statistic $r_{\mbox{\tiny PEL}}(\theta) = \ell_{\mbox{\tiny PEL}}(\theta) - \ell_{\mbox{\tiny PEL}}(\thetaPC)$ at $\theta=\theta^0$ in Theorem \ref{PELWMCCThm}. 
Proofs of Theorem \ref{normalPoint} and Theorem \ref{PELWMCCThm} are given in Section \ref{normalPointproof} and Section \ref{PELWMCCThmproof}, respectively.

\begin{theorem}
\label{normalPoint}
Under certain regularity conditions and a correctly specified propensity score model, we have $\sqrt{n}\left(\thetaPC-\theta^0\right)\stackrel{d}{\rightarrow}\operatorname{N}(0, V_1)$ as $n\rightarrow\infty$, where $V_1=\Sigone^2\Gam^\top\Wone^{-1}\Ome\Wone^{-1}\Gam$ and $\stackrel{d}{\rightarrow}$ denotes convergence in distribution. 
\end{theorem}

In practice, we use ${\hat{\boldsymbol{W}}=\sumzo w_i\sbsji\aij\boldsymbol{g}_{ij}(\thetaPC)\boldsymbol{g}_{ij}(\thetaPC)^\top}$ as an estimate for the limiting matrix $\Wone$, which also leads to $\hat{\sigma}=(\Gam^\top\hat{\boldsymbol{W}}^{-1}\Gam)^{-1}$. 
The estimated variance-covariance matrix ($\boldsymbol{\Omega}$) is $\hat{\boldsymbol{\Omega}}=n^{-1}\sbsj(\hat{\boldsymbol{h}}_j-\bar{\hat{\boldsymbol{h}}})(\hat{\boldsymbol{h}}_j-\bar{\hat{\boldsymbol{h}}})^\top$, where $\hat{\boldsymbol{h}}_j$ is the plug-in estimator of
\[\scalemath{0.76}{
\boldsymbol{h}_j=\left(\begin{array}{c}
0\\
\frac{1}{2}\left\{\left(\frac{\Ri}{\tau_j^0}-1\right)\left(m_{1j}^*-\operatorname{E}\left(m_{1j}^*\right)\right)+\left(\frac{1-\Ri}{1-\tau_j^0}-1\right)\left(m_{0j}^*-\operatorname{E}\left(m_{0j}^*\right)\right)-\left(\boldsymbol{A}+\boldsymbol{E}\right)\boldsymbol{C}^{-1}\bz\left(\Ri-\tau_j^0\right)\right\}\\
\frac{1}{2}\left\{\left(\frac{\Ri}{\tau_j^0}-1\right)\left(m_{1j}^*-\operatorname{E}\left(m_{1j}^*\right)\right)-\left(\frac{1-\Ri}{1-\tau_j^0}-1\right)\left(m_{0j}^*-\operatorname{E}\left(m_{0j}^*\right)\right)-\left(\boldsymbol{A}-\boldsymbol{E}\right)\boldsymbol{C}^{-1}\bz\left(\Ri-\tau^0_j\right)\right\}\\
\frac{\Ri}{\tau_j^0}\left(Y_{1j}-\mu_1^0\right)-\frac{1-\Ri}{1-\tau_j^0}\left(Y_{0j}-\mu_0\right)-\left(\boldsymbol{J}-\boldsymbol{G}\right)\boldsymbol{C}^{-1}\bz\left(\Ri-\tau^0_j\right)
\end{array}\right)}\, 
\]
and $\Bar{\hat{\boldsymbol{h}}}=n^{-1}\sbsj\hat{\boldsymbol{h}}_j$. 
In the above expression of $\boldsymbol{h}_j$, $\tau_j^0=\tau(\bx;\boldsymbol{\alpha}^0)$ is the propensity score under the true propensity score model with $\boldsymbol{\alpha}^0$ representing the true value of $\balpha$,   $m_{ij}^*=m_i\left(\boldsymbol{x}_j, \boldsymbol{\beta}_i^*\right)$ for $i=0,1$, where $\boldsymbol{\beta}_i^*$ is the probability limit of $\hat{\boldsymbol{\beta}}_i$ under the assumed outcome regression model, and
\begin{equation}\scalemath{0.87}{
\begin{aligned}
    \boldsymbol{A}&=-\operatorname{E}\left[\left\{m_{1j}^*-\operatorname{E}(m_{1j}^*)\right\}\left(1-\tau_j^0\right)\bz^\top\right]\, ,\;\;\;
    \boldsymbol{E}=\operatorname{E}\left[\left\{m_{0j}^*-\operatorname{E}(m_{0j}^*)\right\}\tau_j^0\bz^\top\right]\, ,\\
    \boldsymbol{J}&=-\operatorname{E}\left[\Ri\left(Y_{1j}-\mu_1^0\right)\left(1-\tau_j^0\right)\bz^\top/\tau_j^0\right]\, ,\;\;\;
    \boldsymbol{G}=\operatorname{E}\left[\left(1-\Ri\right)\left(Y_{0j}-\mu_0^0\right)\left(1-\tau_j^0\right)^{-1}\tau_j^0\bz^\top\right]\,,\\
    \boldsymbol{C}&=-\operatorname{E}\left[\tau_j^0\left(1-\tau_j^0\right)\bz\bz^\top\right]\, .
\end{aligned}
    \notag}
\end{equation}

Note that the full expression of the model-calibrated PEL ratio function $r_{\mbox{\tiny PEL}}(\theta) = \ell_{\mbox{\tiny PEL}}(\theta) - \ell_{\mbox{\tiny PEL}}(\thetaPC)$ is given by  
\[
    r_{\mbox{\tiny PEL}}(\theta)=\ell_{\mbox{\tiny PEL}}(\hat{\mbox{\boldmath{$p$}}}_1(\theta), \hat{\mbox{\boldmath{$p$}}}_0(\theta))-\ell_{\mbox{\tiny PEL}}\left(\hat{\mbox{\boldmath{$p$}}}_1\left(\thetaPC\right), \hat{\mbox{\boldmath{$p$}}}_0\left(\thetaPC\right)\right)\, .
\]
The asymptotic distribution of $r_{\mbox{\tiny PEL}}(\theta)$ at $\theta=\theta^0$ is given in the theorem below. Note that the rank of the matrix 
$\boldsymbol{M}=\Sigone\boldsymbol{\Omega}^{1/2}\Wone^{-1}\Gam\Gam^\top\Wone^{-1}\boldsymbol{\Omega}^{1/2}$ is one.

\begin{theorem}\label{PELWMCCThm}
Under certain regularity conditions and a correctly specified propensity score model, we have $-2r_{\mbox{\tiny PEL}}(\theta)\stackrel{d}{\rightarrow}\delta\chi^2_1$ when $\theta=\theta^0$, 
where $\delta$ is the unique non-zero eigenvalue of the matrix $\boldsymbol{M}$. 
\end{theorem}

The above result can be used to construct a $100(1-\alpha)\%$ model-calibrated PEL ratio confidence interval for $\theta$ as
\begin{equation}
    \{\theta \; | \; -2r_{\mbox{\tiny PEL}}(\theta)/\hat{\delta}\leq \chi^2_1(\alpha)\}\; ,
    \notag
\end{equation}
where $\hat{\delta}$ is the non-zero eigenvalue of $\hat{\boldsymbol{M}}=\hat{\sigma}\hat{\Ome}^{1/2}\hat{\boldsymbol{W}}^{-1}\Gam\Gam^\top\hat{\boldsymbol{W}}^{-1}\hat{\Ome}^{1/2}$. Computational details and R codes pertaining to the reformulated constrained maximization problem and the method for constructing the model-calibrated PEL ratio confidence interval can be found in \cite{Wu2005}.

\subsection{Bootstrap PEL ratio confidence intervals}
\label{BootstrapPEL}

The bootstrap-calibrated empirical likelihood method can provide an improved approximation to the sampling distribution of the empirical likelihood ratio function, especially when the sample size is small \citep{owen2001}. \cite{Wu2010} proposed a bootstrap procedure for approximating the sampling distribution of the pseudo-empirical likelihood ratio function for complex survey data under certain sampling designs. \cite{Chen2018} discussed bootstrap-calibration procedures for approximating the asymptotic distribution of the PEL ratio function for the nonrandomized pretest-posttest study designs. 

The asymptotic distributions of the proposed PEL ratio functions in Sections \ref{section3} and \ref{section4} involve scaling constants that need to be estimated. Estimation of the scaling constants requires variance estimation, which involves heavy analytic expressions. 
We propose a bootstrap approach to constructing model-calibrated PEL ratio confidence intervals for the ATE as follows. This procedure bypasses the need for estimating the scaling constant required for the $\chi^2$ approximation. 

\begin{description}
    \item[Step 1] Select the $b$-th bootstrap sample $\bs^{[b]}$ of $n$ units from the initial sample $\bs$ by simple random sampling with replacement. Let $\bst^{[b]}=\{j|j\in\bs^{[b]}\text{ and }\Ri=1\}$ and $\bsc^{[b]}=\{j|j\in\bs^{[b]}\text{ and }\Ri=0\}$.   
      
    \item[Step 2] Fit the propensity score model and the outcome regression models to the $b$-th bootstrap sample, and obtain estimates $\hat{\boldsymbol{\alpha}}^{[b]}$, $\hat{\boldsymbol{\beta}}_1^{[b]}$ and $\hat{\boldsymbol{\beta}}_0^{[b]}$.
    Calculate the estimated propensity scores by $\hat{\tau}_j^{[b]}=\tau(\bx; \hat{\boldsymbol{\alpha}}^{[b]})$ and the fitted values $\hat{m}_{1j}^{[b]}=m_1(\boldsymbol{x}_j; \hat{\boldsymbol{\beta}}_1^{[b]})$ and $\hat{m}_{0j}^{[b]}=m_0(\boldsymbol{x}_j; \hat{\boldsymbol{\beta}}_0^{[b]})$ for all the units $j\in \bs^{[b]}$.
    
    \item[Step 3] Construct the bootstrap version of the PEL function as 
    \begin{equation}
         \ell^{[b]}_{\mbox{\tiny PEL}}\left(\boldsymbol{p}_{1}, \boldsymbol{p}_{0}\right)= n\left\{w_{1} \sum_{j\in \bst^{[b]}} \tilde{a}_{1 j}^{[b]} \log \left(p_{1 j}\right) +w_{0} \sum_{j\in \bsc^{[b]}} \tilde{a}_{0 j}^{[b]} \log \left(p_{0 j}\right)\right\}\; ,
     \label{jointPELboot}   
    \end{equation}
    where $\tilde{a}_{1j}^{[b]} = (\hat{\tau}_j^{[b]})^{{-1}}/\sum_{l\in \bst^{[b]}}(\hat{\tau}_l^{[b]})^{{-1}}$ for $j\in\bst^{[b]}$ and $\tilde{a}_{0j}^{[b]} = (1-\hat{\tau}_j^{[b]})^{{-1}}/$\\$\sum_{l\in \bsc^{[b]}}(1-\hat{\tau}_l^{[b]})^{{-1}}$ for $j\in\bsc^{[b]}$.
    Specify the bootstrap versions of the constraints of (\ref{jnorm}), (\ref{para}) and (\ref{jcal}) as 
    \begin{equation}
        \begin{aligned}
        & \sum_{j\in\bst^{[b]}}p_{1j}=1  \;\;\;{\rm and} \;\;\; \sum_{j\in\bsc^{[b]}}p_{0j}=1\, ;\\
        &     \sum_{j\in \bst^{[b]}}p_{1j}Y_{1j}-\sum_{j\in \bsc^{[b]}}p_{0j}Y_{0j}=\theta\,     ; \\
        & \sum_{j \in \bst^{[b]}} p_{1j}\hat{m}_{1j}^{[b]}=\frac{1}{n}\sum_{j\in \bs^{[b]} }\hat{m}_{1j}^{[b]}     \;\;\;{\rm and} \;\;\;    \sum_{j \in \bsc^{[b]}} p_{0j}\hat{m}_{0j}^{[b]}=\frac{1}{n}\sum_{j\in \bs^{[b]}}\hat{m}_{0j}^{[b]}\, .
        \end{aligned}
        \label{constraintsB}
    \end{equation}
    Compute the PEL ratio function $r^{[b]}_{\mbox{\tiny PEL}}(\theta)$ using  $\ell^{[b]}_{\mbox{\tiny PEL}}\left(\boldsymbol{p}_{1}, \boldsymbol{p}_{0}\right)$ with constraints (\ref{constraintsB}) and $\theta=\thetaPC$.
    
   \item[Step 4]  Repeat Steps 1-3 a large number $B$ times, independently, to obtain values of $r_{\mbox{\tiny PEL}}^{[b]}(\theta)$ for $b=1, \dots, B$, all at $\theta=\thetaPC$.
   
\end{description}

The number $B$ is typically set to be $1,000$ but can be larger if more computing resources are available. Let $\alpha^B$ denote the lower $100\alpha$-th quantile of the sequence $r_{\mbox{\tiny PEL}}^{[b]}(\thetaPC)$, $b=1,\dots, B$.
The bootstrap-calibrated $(1-\alpha)$-level PEL ratio confidence interval for $\theta$ is computed as
\[
    \left\{\theta \mid r_{\mbox{\tiny PEL}}(\theta)>{\alpha}^B\right\}\, .
\]

The validity of the bootstrap approach, given that the propensity score model is correctly specified, is justified by deriving the asymptotic distribution of the statistic $-2r^{[b]}_{\mbox{\tiny PEL}}(\theta)$ at $\theta=\thetaPC$. In fact, one can show that the statistic $-2r^{[b]}_{\mbox{\tiny PEL}}(\thetaPC)$ asymptotically follows a scaled $\chi^2_1$ distribution where the scaling constant $\delta^{[b]}$ converges in probability to $\delta$ in Theorem \ref{PELWMCCThm}.
When the propensity score model is misspecified, the proofs of Theorem \ref{normalPoint} and Theorem \ref{PELWMCCThm} are no longer applicable to the bootstrap version. However, the bootstrap procedure still works when the outcome regression models are correctly specified. 
A detailed justification of the bootstrap method is given in Section \ref{validitybootstrap}.

\section{Simulation Studies}
\label{section5}

\subsection{Point estimators and confidence intervals}
\label{simCP}

We conduct limited simulation studies to investigate the performance of our proposed maximum PEL point estimators and PEL ratio confidence intervals for the average treatment effect based on $n_{sim} =1,000$ simulation samples. The results are compared with the commonly used IPW estimator and the AIPW estimator. We use $B=1,000$ for the bootstrap methods with each simulated sample. 
The performance measurement metrics and the candidate methods are as follows.
\begin{itemize}
\item[(i)] For point estimators, we compute the percentage relative bias ($\%RB$) and the mean squared error ($MSE$) for an estimator $\hat\theta$ of the parameter $\theta$ as 
    \[
    \%RB=\frac{1}{n_{sim}}\sum_{s=1}^{n_{sim}}\frac{\hat{\theta}^{(s)}-\theta^0}{\theta^0}\times100\;\;\;{\rm and} \;\;\; MSE=\frac{1}{n_{{sim}}} \sum_{s=1}^{n_{{sim}}}\left(\hat{\theta}^{(s)}-\theta^0\right)^{2}\,,
    \]
    where $\hat{\theta}^{(s)}$ is the estimator $\hat\theta$ computed from the $s$th simulation sample. Results of $\%RB$ and $MSE$ for estimators 
    $\thetaH$, $\thetaP$, $\thetaDRT$, and $\thetaPC$ are presented. 
\item[(ii)] For confidence intervals, we compute the percentage coverage probability ($\%CP$) and the average length ($AL$) for the confidence interval $\mathcal{I}$ as 
\[
\scalemath{0.91}{
\%CP=\frac{1}{n_{sim}}\sum_{s=1}^{n_{sim}}I\left(\theta^0\in \mathcal{I}^{(s)}\right)\times 100} \;\;{\rm and} \;\; \scalemath{0.93}{AL=\frac{1}{n_{sim}}\sum_{s=1}^{n_{sim}} \left(UB^{(s)}-LB^{(s)}\right)}\,,
\]
where $I(\cdot)$ is the indicator function and $\mathcal{I}^{(s)}$ is the interval $\mathcal{I}$ computed from the $s$th simulation sample with $UB^{(s)}$ as the upper bound and $LB^{(s)}$ as the lower bound, i.e., $\mathcal{I}^{(s)} = (LB^{(s)},UB^{(s)})$. We include results of $\%CP$ and $AL$ for the Wald-type confidence interval using $\thetaH$ and the sandwich variance estimator ($\mathcal{I}_{\mbox{\tiny IPW2}}$), the PEL ratio confidence interval based on $\thetaP$ ($\mathcal{I}_{\mbox{\tiny PELR}}$), the Wald-type confidence interval using $\thetaDRT$ and the sandwich variance estimator ($\mathcal{I}_{\mbox{\tiny AIPW2}}$) or the Bootstrap variance estimator ($\mathcal{I}_{\mbox{\tiny AIPW2B}}$),  the model-calibrated PEL ratio confidence interval based on $\thetaPC$ and the scaled chi-squared distribution ($\mathcal{I}_{\mbox{\tiny MCP}}$) or the bootstrap method ($\mathcal{I}_{\mbox{\tiny MCPB}}$). The nominal value of the coverage probability is $95\%$. 
\end{itemize}
 
We consider an independent sample $\bs$ of size $n$ from an infinite population $(Y_1,Y_0,T, \bfx)$, where $\bfx=(x_1,x_2,x_3)^\top$. We consider $n=100$, $200$ and $400$ in the simulation. 
For each subject $j$, $j=1, \dots, n$, $\bx=(x_{j1},x_{j2},x_{j3})^\top$, and the three covariates 
 are generated by $x_{j1}=v_{j1}$, $x_{j2}=v_{j2}+0.2x_{j1}$ and $x_{j3}=v_{j3}+0.3(x_{j1}+x_{j2})$, with $v_{j1}\sim \operatorname{N}(0,1)$, $v_{j2}\sim \operatorname{Bernoulli}(0.6)$ and $v_{j3}\sim \operatorname{Exponential}(1)$. 
The true propensity scores $\tau_j^0 = P(T_j=1 \mid \bfx_j)$ for treatment assignments are given by
\[
\tau:\;\;\;\tau_j^0 = \operatorname{expit}(\alpha_0+0.2x_{j1} +0.2x_{j2}-0.5x_{j3})\, , \;\; j=1, \dots, n, 
\] 
where the value of $\alpha_0$ controls the expected proportion of treatment subjects denoted by $t$. We consider three scenarios with $t=0.3$, $0.5$, or $0.7$. Under the scenario $t=0.3$, for instance, there are approximately $30\%$ of the subjects in the sample $\bs$ belonging to the treatment group. 
The treatment assignment for subject $j$ is decided based on the treatment indicator $\Ri\sim \operatorname{Bernoulli}(\tau_j^0)$.
The outcome regression models are specified as  
\[
m_1:\;\;\;Y_{1j}=4.5+x_{j1}-2x_{j2}+3x_{j3}+a_1\epsilon_j
\] 
 and
\[
m_0:\;\;\;Y_{0j}=1+x_{j1}+x_{j2}+2x_{j3}+a_0\epsilon_j\; ,
\]
where $\epsilon_j\sim \operatorname{N}(0,1)$, $j=1, \dots, n$. The values of $a_1$ and $a_0$ are chosen to control the correlation coefficient $\rho$ between the linear predictor of $\bx$ and the potential outcomes $Y_{1j}$ and $Y_{0j}$.  We consider three scenarios with $\rho=0.3$, $0.5$, or $0.7$, representing weak, mild and strong prediction power of the covariates. For all three scenarios, the true ATE is $\theta^0=2.88$. The sample dataset is given by  $\{(\bx, \Ri, Y_j), j\in \bs\}$, where $Y_j=\Ri Y_{1j}+(1-\Ri)Y_{0j}$. 

We consider three scenarios for the misspecification of the propensity score model $\tau$ and the outcome regression models $m_1$ and $m_0$. 
\begin{itemize}
    \item[(i)] TT: Both the propensity score model and the outcome regression models are correctly specified.
    \item[(ii)] TF: The propensity score model is correctly specified, but the two outcome regression models are misspecified by excluding $x_{3}$ from the model.
    \item[(iii)] FT: The outcome regression models are correctly specified, but the propensity score model is misspecified by omitting $x_3$ in the model. 
\end{itemize}

Our simulation studies are conducted for each of the combinations of $\rho$, $t$, $n$ and working models, resulting in a total of $3\times 3 \times 3 \times 3= 81$ simulation settings.
Tables \ref{PEt03}, \ref{PEt05} and \ref{PEt07} present the percentage relative biases ($\%RB$'s) and the mean squared errors ($MSE$'s)$\times100$ for the point estimators under different combinations of the sample size $n$, the correlation $\rho$, and the model specification scenarios, when the expected treatment proportion $t$ is $0.3$, $0.5$ and $0.7$.
Since the misspecification of the outcome regression models has no impact on $\thetaH$ and $\thetaP$, their simulation results under the ``TF" scenario are identical to the ones under the ``TT" scenario, and hence are not shown.

\begin{table}[h]
    \caption{$\%RB$ and $MSE(\times100)$ for Point Estimators when $t=0.3$}
    \centering
    \begin{tabular}{cclrrrrrr}
    \hline\rule{0pt}{2.1ex}
    \multirow{2}{*}{n}& \multirow{2}{*}{Scenario}&\multirow{2}{*}{Estimator}&\multicolumn{2}{c}{$\rho=0.3$}&\multicolumn{2}{c}{$\rho=0.5$}&\multicolumn{2}{c}{$\rho=0.7$}\\
    \cline{4-9}\rule{0pt}{2.1ex}
    & & &$\% RB$ & $MSE$ &$\% RB$ & $MSE$& $\% RB$ & $MSE$\\
    \hline\rule{0pt}{2.4ex}
        \multirow{10}{*}{100}&\multirow{4}{*}{TT}&$\thetaH$&-6.9 & 635.6 & -6.2 & 208.3 & -5.8 & 90.5 \\
    & &$\thetaP$&-6.9 & 635.6 & -6.2 & 208.3 & -5.8 & 90.5 \\
    & &$\thetaDRT$&-1.4 & 676.0   & -0.8 & 202.2 & -0.5 & 71.8 \\
    & &$\thetaPC$&-1.6 & 686.4 & -1.0   & 204.8 & -0.6 & 72.6\\
    \cline{2-9}\rule{0pt}{2.4ex}
    &\multirow{2}{*}{TF}&$\thetaDRT$&-8.4 & 624.5 & -7.8 & 203.0   & -7.5 & 87.0   \\
    & &$\thetaPC$&-9.6 & 609.6 & -8.8 & 195.7 & -8.4 & 82.4 \\
    \cline{2-9}\rule{0pt}{2.4ex}
    &\multirow{4}{*}{FT}&$\thetaH$&-38.2 & 680.4 & -37.4 & 299.6 & -37.0  & 193.5 \\
    & &$\thetaP$&-38.2 & 680.4 & -37.4 & 299.6 & -37.0  & 193.5 \\
    & &$\thetaDRT$&-1.3  & 665.3 & -0.7  & 199.1 & -0.4 & 70.8  \\
    & &$\thetaPC$&-1.1  & 694.9 & -0.8  & 208.6 & -0.5 & 74.1 \\
    \hline\rule{0pt}{2.4ex}
    \multirow{10}{*}{200}&\multirow{4}{*}{TT}&$\thetaH$&-5.0   & 334.4 & -4.0   & 118.7 & -3.6 & 61.7 \\
    & &$\thetaP$&-5.0   & 334.4 & -4.0   & 118.7 & -3.6 & 61.7 \\
    & &$\thetaDRT$&-2.1 & 327.1 & -1.2 & 98.6  & -0.8 & 35.4 \\
    & &$\thetaPC$&-1.9 & 327.8 & -1.1 & 98.8  & -0.8 & 35.4\\
    \cline{2-9}\rule{0pt}{2.4ex}
    &\multirow{2}{*}{TF}&$\thetaH$&-5.8 & 327.8 & -4.8 & 114.6 & -4.3 & 59.4 \\
    & &$\thetaPC$&-6.5 & 323.1 & -5.7 & 108.1 & -5.0   & 53.1 \\
    \cline{2-9}\rule{0pt}{2.4ex}
    &\multirow{4}{*}{FT}&$\thetaH$&-38.0  & 410.1 & -37.3 & 210.3 & -37.0  & 154.3 \\
    & &$\thetaP$&-38.0  & 410.1 & -37.3 & 210.3 & -37.0  & 154.3 \\
    & &$\thetaDRT$&-2.1 & 324.5 & -1.3  & 97.8  & -0.8 & 35.1  \\
    & &$\thetaPC$&-2.4 & 329.8 & -1.5  & 100.2 & -1.0   & 36.0  \\
    \hline\rule{0pt}{2.4ex}
    \multirow{10}{*}{400}&\multirow{4}{*}{TT}&$\thetaH$&-3.8 & 168.8 & -2.7 & 58.1 & -2.2 & 27.7 \\
    & &$\thetaP$&-3.8 & 168.8 & -2.7 & 58.1 & -2.2 & 27.7 \\
    & &$\thetaDRT$&-2.3 & 158.0   & -1.3 & 47.3 & -0.7 & 16.8 \\
    & &$\thetaPC$&-2.3 & 156.8 & -1.2 & 46.9 & -0.7 & 16.7\\
    \cline{2-9}\rule{0pt}{2.4ex}
    &\multirow{2}{*}{TF}&$\thetaDRT$&-4.1 & 166.3 & -3.0   & 56.6 & -2.5 & 26.6 \\
    & &$\thetaPC$&-4.4 & 164.8 & -3.3 & 55.7 & -2.8 & 25.7 \\
    \cline{2-9}\rule{0pt}{2.4ex}
    &\multirow{4}{*}{FT}&$\thetaH$&-38.2 & 262.1 & -37.4 & 161.5 & -36.9 & 132.7 \\
    & &$\thetaP$&-38.2 & 262.1 & -37.4 & 161.5 & -36.9 & 132.7 \\
    & &$\thetaDRT$&-1.9  & 153.4 & -1.1  & 45.9  & -0.6  & 16.3  \\
    & &$\thetaPC$&-1.9  & 157.9 & -1.1  & 47.3  & -0.6  & 16.8 \\
    \hline
    \end{tabular}
    \label{PEt03}
\end{table}

\begin{table}[h]
    \caption{$\%RB$ and $MSE(\times100)$ for Point Estimators when $t=0.5$}
    \centering
    \begin{tabular}{cclrrrrrr}
    \hline\rule{0pt}{2.1ex}
    \multirow{2}{*}{n}& \multirow{2}{*}{Scenario}&\multirow{2}{*}{Estimator}&\multicolumn{2}{c}{$\rho=0.3$}&\multicolumn{2}{c}{$\rho=0.5$}&\multicolumn{2}{c}{$\rho=0.7$}\\
    \cline{4-9}\rule{0pt}{2.1ex}
    & & &$\% RB$ & $MSE$ &$\% RB$ & $MSE$& $\% RB$ & $MSE$\\
    \hline\rule{0pt}{2.4ex}
        \multirow{10}{*}{100}&\multirow{4}{*}{TT}&$\thetaH$&0.1 & 483.9 & -1.1 & 164.5 & -1.7 & 75.0   \\
    & &$\thetaP$&0.1 & 483.9 & -1.1 & 164.5 & -1.7 & 75.0   \\
    & &$\thetaDRT$&2.2 & 459.0   & 1.2  & 139.2 & 0.7  & 50.7 \\
    & &$\thetaPC$&2.2 & 453.7 & 1.2  & 137.6 & 0.7  & 50.2\\
    \cline{2-9}\rule{0pt}{2.4ex}
    &\multirow{2}{*}{TF}&$\thetaDRT$&-0.8 & 474.5 & -2.0   & 156.3 & -2.6 & 67.2 \\
    & &$\thetaPC$&-1.3 & 461.8 & -2.5 & 147.9 & -3.0   & 60.4 \\
    \cline{2-9}\rule{0pt}{2.4ex}
    &\multirow{4}{*}{FT}&$\thetaH$&-34.7 & 513.7 & -36.2 & 245.6 & -37.0 & 174.2 \\
    & &$\thetaP$&-34.7 & 513.7 & -36.2 & 245.6 & -37.0 & 174.2 \\
    & &$\thetaDRT$&2.7   & 451.3 & 1.5   & 136.7 & 0.8 & 49.8  \\
    & &$\thetaPC$&2.2   & 451.0   & 1.2   & 136.5 & 0.7 & 49.6 \\
    \hline\rule{0pt}{2.4ex}
    \multirow{10}{*}{200}&\multirow{4}{*}{TT}&$\thetaH$&0.6 & 226.4 & -0.7 & 72.0   & -1.3 & 29.6 \\
    & &$\thetaP$&0.6 & 226.4 & -0.7 & 72.0   & -1.3 & 29.6 \\
    & &$\thetaDRT$&2.7 & 224.2 & 1.4  & 67.6 & 0.7  & 24.4 \\
    & &$\thetaPC$&2.7 & 224.1 & 1.3  & 67.5 & 0.7  & 24.4\\
    \cline{2-9}\rule{0pt}{2.4ex}
    &\multirow{2}{*}{TF}&$\thetaDRT$&0.2  & 224.9 & -1.1 & 71.2 & -1.7 & 29.2 \\
    & &$\thetaPC$&-0.1 & 224.1 & -1.3 & 70.7 & -1.9 & 28.7 \\
    \cline{2-9}\rule{0pt}{2.4ex}
    &\multirow{4}{*}{FT}&$\thetaH$&-35.8 & 326.1 & -37.2 & 188.2 & -37.9 & 152.2 \\
    & &$\thetaP$&-35.8 & 326.1 & -37.2 & 188.2 & -37.9 & 152.2 \\
    & &$\thetaDRT$&2.6   & 221.9 & 1.3   & 66.9  & 0.7   & 24.1  \\
    & &$\thetaPC$&2.5   & 222.8 & 1.2   & 67.1  & 0.6   & 24.2 \\
    \hline\rule{0pt}{2.4ex}
    \multirow{10}{*}{400}&\multirow{4}{*}{TT}&$\thetaH$&-1.3 & 122.5 & -1.1 & 38.3 & -1.0   & 15.3 \\
    & &$\thetaP$&-1.3 & 122.5 & -1.1 & 38.3 & -1.0   & 15.3 \\
    & &$\thetaDRT$&-0.4 & 122.2 & -0.2 & 37.0   & -0.1 & 13.4 \\
    & &$\thetaPC$&-0.4 & 121.5 & -0.2 & 36.8 & -0.1 & 13.3\\
    \cline{2-9}\rule{0pt}{2.4ex}
    &\multirow{2}{*}{TF}&$\thetaDRT$&-1.6 & 122.6 & -1.4 & 38.4 & -1.3 & 15.3 \\
    & &$\thetaPC$&-1.7 & 122.4 & -1.5 & 38.3 & -1.4 & 15.2 \\
    \cline{2-9}\rule{0pt}{2.4ex}
    &\multirow{4}{*}{FT}&$\thetaH$&-37.8 & 234.8 & -37.6 & 155.9 & -37.5 & 133.9 \\
    & &$\thetaP$&-37.8 & 234.8 & -37.6 & 155.9 & -37.5 & 133.9 \\
    & &$\thetaDRT$&-0.5  & 119.4 & -0.3  & 36.2  & -0.2  & 13.1  \\
    & &$\thetaPC$&-0.5  & 119.6 & -0.3  & 36.3  & -0.2  & 13.2\\
    \hline
    \end{tabular}
    \label{PEt05}
\end{table}

\begin{table}[h]
    \caption{$\%RB$ and $MSE(\times100)$ for Point Estimators when $t=0.7$}
    \centering
    \begin{tabular}{cclrrrrrr}
    \hline\rule{0pt}{2.1ex}
    \multirow{2}{*}{n}& \multirow{2}{*}{Scenario}&\multirow{2}{*}{Estimator}&\multicolumn{2}{c}{$\rho=0.3$}&\multicolumn{2}{c}{$\rho=0.5$}&\multicolumn{2}{c}{$\rho=0.7$}\\
    \cline{4-9}\rule{0pt}{2.1ex}
    & & &$\% RB$ & $MSE$ &$\% RB$ & $MSE$& $\% RB$ & $MSE$\\
    \hline\rule{0pt}{2.4ex}
        \multirow{10}{*}{100}&\multirow{4}{*}{TT}&$\thetaH$&-1.4 & 544.9 & -1.3 & 168.0   & -1.3 & 64.2 \\
    & &$\thetaP$&-1.4 & 544.9 & -1.3 & 168.0   & -1.3 & 64.2 \\
    & &$\thetaDRT$&-0.6 & 545.9 & -0.3 & 164.7 & -0.2 & 59.4 \\
    & &$\thetaPC$&-0.6 & 546.3 & -0.3 & 164.6 & -0.2 & 59.3\\
    \cline{2-9}\rule{0pt}{2.4ex}
    &\multirow{2}{*}{TF}&$\thetaDRT$&-2.6 & 543.1 & -2.5 & 165.7 & -2.4 & 61.9 \\
    & &$\thetaPC$&-2.7 & 543.0   & -2.6 & 165.7 & -2.6 & 61.9 \\
    \cline{2-9}\rule{0pt}{2.4ex}
    &\multirow{4}{*}{FT}&$\thetaH$&-40.9 & 648.6 & -40.8 & 312.2 & -40.8 & 219.5 \\
    & &$\thetaP$&-40.9 & 648.6 & -40.8 & 312.2 & -40.8 & 219.5 \\
    & &$\thetaDRT$&-0.8  & 531.5 & -0.4  & 160.4 & -0.3  & 57.9  \\
    & &$\thetaPC$&-0.7  & 540.2 & -0.2  & 162.2 & -0.1  & 58.3 \\
    \hline\rule{0pt}{2.4ex}
    \multirow{10}{*}{200}&\multirow{4}{*}{TT}&$\thetaH$&0.7 & 259.3 & -0.2 & 78.5 & -0.6 & 29.1 \\
    & &$\thetaP$&0.7 & 259.3 & -0.2 & 78.5 & -0.6 & 29.1 \\
    & &$\thetaDRT$&1.6 & 256.9 & 0.8  & 76.5 & 0.3  & 27.1 \\
    & &$\thetaPC$&1.7 & 256.5 & 0.8  & 76.4 & 0.4  & 27.0  \\
    \cline{2-9}\rule{0pt}{2.4ex}
    &\multirow{2}{*}{TF}&$\thetaDRT$&0.0 & 257.8 & -0.9 & 77.9 & -1.3 & 28.6 \\
    & &$\thetaPC$&0.0 & 257.5 & -0.9 & 77.8 & -1.3 & 28.5 \\
    \cline{2-9}\rule{0pt}{2.4ex}
    &\multirow{4}{*}{FT}&$\thetaH$&-39.7 & 383.8 & -40.7 & 226.0  & -41.2 & 183.0  \\
    & &$\thetaP$&-39.7 & 383.8 & -40.7 & 226.0  & -41.2 & 183.0  \\
    & &$\thetaDRT$&1.5   & 251.3 & 0.7   & 74.8 & 0.3   & 26.5 \\
    & &$\thetaPC$&1.4   & 251.7 & 0.7   & 74.9 & 0.3   & 26.5\\
    \hline\rule{0pt}{2.4ex}
    \multirow{10}{*}{400}&\multirow{4}{*}{TT}&$\thetaH$&2.3 & 129.2 & 1.2 & 40.3 & 0.6 & 15.8 \\
    & &$\thetaP$&2.3 & 129.2 & 1.2 & 40.3 & 0.6 & 15.8 \\
    & &$\thetaDRT$&2.6 & 126.1 & 1.4 & 38.0   & 0.8 & 13.7 \\
    & &$\thetaPC$&2.6 & 125.8 & 1.4 & 37.9 & 0.8 & 13.6\\
    \cline{2-9}\rule{0pt}{2.4ex}
    &\multirow{2}{*}{TF}&$\thetaDRT$&2.1 & 128.5 & 0.9 & 39.9 & 0.3 & 15.4 \\
    & &$\thetaPC$&2.0   & 128.2 & 0.9 & 39.7 & 0.3 & 15.3 \\
    \cline{2-9}\rule{0pt}{2.4ex}
    &\multirow{4}{*}{FT}&$\thetaH$&-38.5 & 245.3 & -39.6 & 171.6 & -40.1 & 152.9 \\
    & &$\thetaP$&-38.5 & 245.3 & -39.6 & 171.6 & -40.1 & 152.9 \\
    & &$\thetaDRT$&2.5   & 122.8 & 1.4   & 37.0    & 0.8   & 13.3  \\
    & &$\thetaPC$&2.5   & 124.0   & 1.3   & 37.3  & 0.8   & 13.4 \\
    \hline
    \end{tabular}
    \label{PEt07}
\end{table}

From the simulation results, we can see that (1) $\thetaH$ and $\thetaP$ are exactly the same, and they perform well when the propensity score model is correctly specified (``TT"); although the approximately $-6\%$ $RB$ is slightly large in the case of $(n, t)=(100, 0.3)$, this might be due to the small sample size in the treatment group; there are noticeable biases when the propensity score model is not correctly specified (``FT"), revealing the failure of the two estimators in this scenario; (2) the performances of $\thetaDRT$ and $\thetaPC$ are similar as expected, and they perform well in all of the three model specification scenarios, showing that they are doubly robust; although the absolute values of the $RB$'s of $\thetaDRT$ and $\thetaPC$ are over $5\%$ for some ``TF" cases, these values are still reasonable because those of $\thetaH$ and $\thetaP$ are close to or over $5\%$ in the corresponding ``TT" cases (for example, $(n,t)=(100, 0.3)$) and the consistency of $\thetaDRT$ or $\thetaPC$ relies on the propensity score model in ``TF" cases; (3) when both models are correctly specified (``TT"), the mean squared errors of $\thetaDRT$ and $\thetaPC$ seem to be smaller than those of $\thetaH$ and $\thetaP$; and (4) the $MSE$ of each estimator decreases as $n$ increases.

The percentage coverage probabilities ($\%CP$'s) and average lengths ($AL$'s) $\times 100$ of the $95\%$ confidence intervals under different settings are reported in Tables \ref{CIt03}, \ref{CIt05} and \ref{CIt07} for $t=0.3$, $t=0.5$ and $t=0.7$, respectively. 
Since $\mathcal{I}_{\mbox{\tiny IPW2}}$ and $\mathcal{I}_{\mbox{\tiny PELR}}$ do not rely on the outcome regression models, their simulation results in ``TT" and ``TF" are identical and are only presented for the scenario ``TT". 
The results show that (1) when both models are correctly specified (``TT"), none of these methods fails; 
(2) all methods work well if the outcome regression models are misspecified but the propensity score model is correctly specified (``TF"); 
(3) the two confidence intervals $\mathcal{I}_{\mbox{\tiny IPW2}}$ and $\mathcal{I}_{\mbox{\tiny PELR}}$ are not reliable when the propensity score model is misspecified (``FT"), as the corresponding point estimators are not valid in this scenario;
(4) confidence intervals $\mathcal{I}_{\mbox{\tiny AIPW2}}$, $\mathcal{I}_{\mbox{\tiny AIPW2B}}$, $\mathcal{I}_{\mbox{\tiny MCP}}$ and $\mathcal{I}_{\mbox{\tiny MCPB}}$ seem to perform well in all three model specification scenarios, meaning that they are doubly robust to the misspecification of either the propensity score model or the outcome regression models;
(5) the intervals $\mathcal{I}_{\mbox{\tiny AIPW2B}}$ and $\mathcal{I}_{\mbox{\tiny MCPB}}$ are usually wider than the other two with higher coverage probabilities, which is an advantage of the bootstrap-calibrated methods in some settings (such as $(t, n)=(0.3, 200)$) where the coverage probabilities of the other two are close to $92\%$, slightly low compared to the nominal level of $95\%$ ;  
(6) between the two bootstrap-calibrated methods, $\mathcal{I}_{\mbox{\tiny MCPB}}$ is usually wider than $\mathcal{I}_{\mbox{\tiny AIPW2B}}$ with higher coverage probabilities; 
(7) when the sample size is small, for scenario ``TT", as $\rho$ increases, the coverage probability of $\mathcal{I}_{\mbox{\tiny MCP}}$ reaches to the nominal level of $95\%$ faster than $\mathcal{I}_{\mbox{\tiny AIPW2}}$; for example, when $\rho=0.7$ and $(n, t)=(100, 0.3)$, the coverage probability of $\mathcal{I}_{\mbox{\tiny MCP}}$ is $95.2\%$, while that of $\mathcal{I}_{\mbox{\tiny AIPW2}}$ is $92.7\%$; however, in other scenarios (``TF" and ``FT"), this advantage disappears;
(8) all confidence intervals based on doubly robust methods are similar to each other and perform well when $n$ is large and get narrower as $n$ increases.

\begin{table}[pt]
    \caption{$\%CP$ and $AL(\times100)$ for $95\%$ Confidence Intervals when $t=0.3$}
    \centering
    \begin{tabular}{cclrrrrrr}
    \hline\rule{0pt}{2.1ex}
    \multirow{2}{*}{n}& \multirow{2}{*}{Scenario}&\multirow{2}{*}{Estimator}&\multicolumn{2}{c}{$\rho=0.3$}&\multicolumn{2}{c}{$\rho=0.5$}&\multicolumn{2}{c}{$\rho=0.7$}\\
    \cline{4-9}\rule{0pt}{2.1ex}
    & & &$\% CP$ & $AL$ &$\% CP$ & $AL$& $\% CP$ & $AL$\\
    \hline\rule{0pt}{2.4ex}
    \multirow{16}{*}{100}&\multirow{6}{*}{TT}&$\mathcal{I}_{\mbox{\tiny IPW2}}$&93.4 & 916.6  & 93.6 & 518.2 & 92.4 & 331.2 \\
    & &$\mathcal{I}_{\mbox{\tiny PELR}}$&93.9 & 926.0    & 93.7 & 522.6 & 92.7 & 333.3 \\
    & &$\mathcal{I}_{\mbox{\tiny AIPW2}}$&92.1 & 938.2 & 92.4 & 514.0   & 92.7 & 307.5 \\
    & &$\mathcal{I}_{\mbox{\tiny AIPW2B}}$&95.5 & 1109.7 & 96.1 & 607.0   & 96.2 & 361.5 \\
    & &$\mathcal{I}_{\mbox{\tiny MCP}}$&93.0   & 974.7 & 93.5 & 570.8 & 95.2 & 386.5 \\
    & &$\mathcal{I}_{\mbox{\tiny MCPB}}$&97.0   & 1241.0   & 96.9 & 677.4 & 97.2 & 404.0 \\
    \cline{2-9}\rule{0pt}{2.1ex}
    &\multirow{4}{*}{TF}&$\mathcal{I}_{\mbox{\tiny AIPW2}}$&93.7 & 915.8 & 93.0   & 515.2 & 92.1 & 325.8 \\
    & &$\mathcal{I}_{\mbox{\tiny AIPW2B}}$&95.8 & 1005.1 & 95.8 & 568.5 & 94.7 & 365.0   \\
    & &$\mathcal{I}_{\mbox{\tiny MCP}}$&93.6 & 933.0   & 92.6 & 529.1 & 91.4 & 340.0   \\
    & &$\mathcal{I}_{\mbox{\tiny MCPB}}$&96.4 & 1066.8 & 96.2 & 607.2 & 95.3 & 396.4 \\
    \cline{2-9}\rule{0pt}{2.1ex}
    &\multirow{6}{*}{FT}&$\mathcal{I}_{\mbox{\tiny IPW2}}$&91.1 & 902.0    & 85.8 & 516.3 & 75.5 & 340.3 \\
    & &$\mathcal{I}_{\mbox{\tiny PELR}}$&91.5 & 910.7  & 86.1 & 521.3 & 75.2 & 343.8 \\
    & &$\mathcal{I}_{\mbox{\tiny AIPW2}}$&92.4 & 950.3 & 93.0   & 520.6 & 92.9 & 311.2 \\
    & &$\mathcal{I}_{\mbox{\tiny AIPW2B}}$&95.5 & 1098.2 & 95.5 & 600.8 & 95.5 & 357.9 \\
    & &$\mathcal{I}_{\mbox{\tiny MCP}}$&90.9 & 895.6 & 90.5 & 493.7 & 92.2 & 301.1 \\
    & &$\mathcal{I}_{\mbox{\tiny MCPB}}$&96.9 & 1249.6 & 96.8 & 683.2 & 97.0   & 407.3\\
    \hline\rule{0pt}{2.1ex}
    \multirow{16}{*}{200}&\multirow{6}{*}{TT}&$\mathcal{I}_{\mbox{\tiny IPW2}}$&91.6 & 664.2 & 92.0   & 378.1 & 90.9 & 243.9 \\
    & &$\mathcal{I}_{\mbox{\tiny PELR}}$&91.6 & 668.3 & 92.1 & 380.3 & 91.1 & 245.2 \\
    & &$\mathcal{I}_{\mbox{\tiny AIPW2}}$&91.9 & 667.2 & 92.0   & 365.5 & 91.9 & 218.5 \\
    & &$\mathcal{I}_{\mbox{\tiny AIPW2B}}$&93.8 & 714.8 & 93.8 & 391.3 & 94.0   & 233.5 \\
    & &$\mathcal{I}_{\mbox{\tiny MCP}}$&91.6 & 693.8 & 92.4 & 402.8 & 94.6 & 269.4 \\
    & &$\mathcal{I}_{\mbox{\tiny MCPB}}$&94.7 & 750.1 & 94.4 & 411.0   & 94.7 & 244.9\\
    \cline{2-9}\rule{0pt}{2.1ex}
    &\multirow{4}{*}{TF}&$\mathcal{I}_{\mbox{\tiny AIPW2}}$&91.6 & 662.5 & 91.7 & 376.0   & 89.5 & 241.1 \\
    & &$\mathcal{I}_{\mbox{\tiny AIPW2B}}$&93.0   & 697.9 & 92.8 & 400.6 & 91.9 & 262.9 \\
    & &$\mathcal{I}_{\mbox{\tiny MCP}}$&91.8 & 680.4 & 91.5 & 387.0   & 89.3 & 250.2 \\
    & &$\mathcal{I}_{\mbox{\tiny MCPB}}$&93.9 & 713.9 & 93.3 & 410.1 & 92.9 & 273.0   \\
    \cline{2-9}\rule{0pt}{2.1ex}
    &\multirow{6}{*}{FT}&$\mathcal{I}_{\mbox{\tiny IPW2}}$&89.2&647.2&78.3&370.1&60.4&243.6\\
    & &$\mathcal{I}_{\mbox{\tiny PELR}}$&89.3 & 650.9 & 78.4 & 372.2 & 60.7 & 245.0   \\
    & &$\mathcal{I}_{\mbox{\tiny AIPW2}}$&92.2 & 670.4 & 92.4 & 367.2 & 92.5 & 219.5 \\
    & &$\mathcal{I}_{\mbox{\tiny AIPW2B}}$&93.7 & 707.5 & 93.4 & 387.3 & 94.0   & 231.2 \\
    & &$\mathcal{I}_{\mbox{\tiny MCP}}$&91.2 & 644.7 & 91.3 & 355.2 & 91.8 & 214.9 \\
    & &$\mathcal{I}_{\mbox{\tiny MCPB}}$&94.7 & 755.2 & 95.1 & 414.3 & 95.2 & 246.8\\
    \hline\rule{0pt}{2.1ex}
    \multirow{16}{*}{400}&\multirow{6}{*}{TT}&$\mathcal{I}_{\mbox{\tiny IPW2}}$&94.0   & 478.8 & 93.9 & 272.3 & 93.3 & 176.1 \\
    & &$\mathcal{I}_{\mbox{\tiny PELR}}$&94.1 & 480.4 & 93.9 & 273.2 & 93.3 & 176.7 \\
    & &$\mathcal{I}_{\mbox{\tiny AIPW2}}$&94.0   & 473.2 & 94.3 & 259.2 & 94.4 & 155.0   \\
    & &$\mathcal{I}_{\mbox{\tiny AIPW2B}}$&95.3 & 488.9 & 95.3 & 267.7 & 95.2 & 159.9 \\
    & &$\mathcal{I}_{\mbox{\tiny MCP}}$&94.2 & 481.6 & 94.8 & 276.3 & 95.9 & 180.7 \\
    & &$\mathcal{I}_{\mbox{\tiny MCPB}}$&95.3 & 499.0   & 96.2 & 273.1 & 95.7 & 163.1\\
    \cline{2-9}\rule{0pt}{2.1ex}
    &\multirow{4}{*}{TF}&$\mathcal{I}_{\mbox{\tiny AIPW2}}$&94.0   & 477.3 & 94.3 & 270.5 & 93.3 & 174.0   \\
    & &$\mathcal{I}_{\mbox{\tiny AIPW2B}}$&94.5 & 491.6 & 95.0   & 281.9 & 94.5 & 185.4 \\
    & &$\mathcal{I}_{\mbox{\tiny MCP}}$&94.2 & 479.5 & 94.1 & 272.9 & 93.0   & 176.5 \\
    & &$\mathcal{I}_{\mbox{\tiny MCPB}}$&95.4 & 498.0   & 95.3 & 285.7 & 94.7 & 188.5 \\
    \cline{2-9}\rule{0pt}{2.1ex}
    &\multirow{6}{*}{FT}&$\mathcal{I}_{\mbox{\tiny IPW2}}$&83.8 & 457.4 & 62.6 & 261.7 & 31.4 & 172.3 \\
    & &$\mathcal{I}_{\mbox{\tiny PELR}}$&84.0   & 458.7 & 62.6 & 262.4 & 31.7 & 172.9 \\
    & &$\mathcal{I}_{\mbox{\tiny AIPW2}}$&94.1 & 473.4 & 93.8 & 259.3 & 94.3 & 155.0   \\
    & &$\mathcal{I}_{\mbox{\tiny AIPW2B}}$&95.0   & 483.4 & 94.9 & 264.8 & 94.8 & 158.2 \\
    & &$\mathcal{I}_{\mbox{\tiny MCP}}$&92.8 & 457.0   & 93.4 & 251.6 & 94.0   & 151.8\\
    & &$\mathcal{I}_{\mbox{\tiny MCPB}}$&95.2 & 502.6 & 95.3 & 275.2 & 95.3 & 164.1\\
    \hline
    \label{CIt03}
    \end{tabular}
\end{table}

\begin{table}[pt]
    \caption{$\%CP$ and $AL(\times100)$ for $95\%$ Confidence Intervals when $t=0.5$}
    \centering
    \begin{tabular}{cclrrrrrr}
    \hline\rule{0pt}{2.1ex}
    \multirow{2}{*}{n}& \multirow{2}{*}{Scenario}&\multirow{2}{*}{Estimator}&\multicolumn{2}{c}{$\rho=0.3$}&\multicolumn{2}{c}{$\rho=0.5$}&\multicolumn{2}{c}{$\rho=0.7$}\\
    \cline{4-9}\rule{0pt}{2.1ex}
    & & &$\% CP$ & $AL$ &$\% CP$ & $AL$& $\% CP$ & $AL$\\
    \hline\rule{0pt}{2.4ex}
    \multirow{16}{*}{100}&\multirow{6}{*}{TT}&$\mathcal{I}_{\mbox{\tiny IPW2}}$&94.0   & 819.8 & 93.5 & 459.9 & 93.6 & 288.5 \\
    & &$\mathcal{I}_{\mbox{\tiny PELR}}$&94.2 & 827.1 & 93.7 & 463.2 & 93.6 & 289.6 \\
    & &$\mathcal{I}_{\mbox{\tiny AIPW2}}$&93.6 & 822.0   & 93.7 & 451.0   & 93.4 & 270.7 \\
    & &$\mathcal{I}_{\mbox{\tiny AIPW2B}}$&95.5 & 873.4 & 95.0   & 478.8 & 95.2 & 286.8 \\
    & &$\mathcal{I}_{\mbox{\tiny MCP}}$&94.4 & 843.2 & 95.0   & 479.4 & 95.8 & 311.8 \\
    & &$\mathcal{I}_{\mbox{\tiny MCPB}}$&96.9 & 927.8 & 96.7 & 508.7 & 96.3 & 304.9\\
    \cline{2-9}\rule{0pt}{2.1ex}
    &\multirow{4}{*}{TF}&$\mathcal{I}_{\mbox{\tiny AIPW2}}$&94.2 & 818.8 & 93.5 & 458.0   & 93.4 & 285.6 \\
    & &$\mathcal{I}_{\mbox{\tiny AIPW2B}}$&94.7 & 860.6 & 95.0   & 484.4 & 95.0   & 307.3 \\
    & &$\mathcal{I}_{\mbox{\tiny MCP}}$&94.4 & 828.7 & 93.7 & 463.6 & 93.4 & 290.2 \\
    & &$\mathcal{I}_{\mbox{\tiny MCPB}}$&96.2 & 894.6 & 96.4 & 504.0   & 95.9 & 321.3 \\
    \cline{2-9}\rule{0pt}{2.1ex}
    &\multirow{6}{*}{FT}&$\mathcal{I}_{\mbox{\tiny IPW2}}$&92.9 & 811.2 & 86.5 & 468.1 & 73.4 & 313.4 \\
    & &$\mathcal{I}_{\mbox{\tiny PELR}}$&92.9 & 818.0   & 86.8 & 471.9 & 73.6 & 316.1 \\
    & &$\mathcal{I}_{\mbox{\tiny AIPW2}}$&94.5 & 821.7 & 94.2 & 450.8 & 94.0   & 270.6 \\
    & &$\mathcal{I}_{\mbox{\tiny AIPW2B}}$&95.7 & 862.9 & 95.3 & 473.1 & 95.5 & 283.5 \\
    & &$\mathcal{I}_{\mbox{\tiny MCP}}$&93.4 & 804.9 & 93.8 & 442.4 & 93.5 & 266.8 \\
    & &$\mathcal{I}_{\mbox{\tiny MCPB}}$&97.0   & 924.7 & 96.6 & 508.0   & 96.6 & 304.5\\
    \hline\rule{0pt}{2.1ex}
    \multirow{16}{*}{200}&\multirow{6}{*}{TT}&$\mathcal{I}_{\mbox{\tiny IPW2}}$&94.8 & 583.4 & 94.8 & 326.2 & 94.2 & 204.0   \\
    & &$\mathcal{I}_{\mbox{\tiny PELR}}$&95.2 & 586.1 & 95.2 & 327.6 & 94.4 & 204.9 \\
    & &$\mathcal{I}_{\mbox{\tiny AIPW2}}$&95.2 & 581.4 & 95.5 & 319.1 & 95.5 & 191.7 \\
    & &$\mathcal{I}_{\mbox{\tiny AIPW2B}}$&95.9 & 598.8 & 96.1 & 328.5 & 96.0   & 197.2 \\
    & &$\mathcal{I}_{\mbox{\tiny MCP}}$&95.5 & 587.6 & 95.6 & 331.4 & 96.2 & 211.0   \\
    & &$\mathcal{I}_{\mbox{\tiny MCPB}}$&96.6 & 614.5 & 97.0   & 337.2 & 96.4 & 202.2\\
    \cline{2-9}\rule{0pt}{2.1ex}
    &\multirow{4}{*}{TF}&$\mathcal{I}_{\mbox{\tiny AIPW2}}$&94.6 & 582.3 & 94.7 & 325.2 & 94.1 & 203.0   \\
    & &$\mathcal{I}_{\mbox{\tiny AIPW2B}}$&95.4 & 599.7 & 95.3 & 338.4 & 95.0   & 215.6 \\
    & &$\mathcal{I}_{\mbox{\tiny MCP}}$&94.8 & 584.7 & 94.5 & 326.7 & 93.8 & 204.3 \\
    & &$\mathcal{I}_{\mbox{\tiny MCPB}}$&95.8 & 610.4 & 95.9 & 344.0   & 95.5 & 219.3 \\
    \cline{2-9}\rule{0pt}{2.1ex}
    &\multirow{6}{*}{FT}&$\mathcal{I}_{\mbox{\tiny IPW2}}$&88.8 & 575.6 & 76.6 & 332.2 & 51.4 & 222.4 \\
    & &$\mathcal{I}_{\mbox{\tiny PELR}}$&88.7 & 578.2 & 76.8 & 333.7 & 52.0   & 223.5 \\
    & &$\mathcal{I}_{\mbox{\tiny AIPW2}}$&95.9 & 581.6 & 95.6 & 319.2 & 95.7 & 191.7 \\
    & &$\mathcal{I}_{\mbox{\tiny AIPW2B}}$&96.4 & 593.3 & 96.8 & 325.5 & 96.4 & 195.4 \\
    & &$\mathcal{I}_{\mbox{\tiny MCP}}$&95.1 & 571.7 & 94.8 & 314.4 & 95.2 & 189.3 \\
    & &$\mathcal{I}_{\mbox{\tiny MCPB}}$&96.8 & 614.8 & 96.8 & 337.3 & 96.5 & 202.2\\
    \hline\rule{0pt}{2.1ex}
    \multirow{16}{*}{400}&\multirow{6}{*}{TT}&$\mathcal{I}_{\mbox{\tiny IPW2}}$&92.9 & 413.0   & 92.5 & 230.8 & 92.5 & 144.6 \\
    & &$\mathcal{I}_{\mbox{\tiny PELR}}$&93.0   & 414.0   & 92.7 & 231.3 & 92.5 & 144.9 \\
    & &$\mathcal{I}_{\mbox{\tiny AIPW2}}$&92.7 & 410.9 & 92.7 & 225.5 & 92.7 & 135.4 \\
    & &$\mathcal{I}_{\mbox{\tiny AIPW2B}}$&93.4 & 416.3 & 93.0   & 228.4 & 93.0   & 137.1 \\
    & &$\mathcal{I}_{\mbox{\tiny MCP}}$&92.9 & 412.4 & 93.2 & 229.4 & 93.5 & 142.2 \\
    & &$\mathcal{I}_{\mbox{\tiny MCPB}}$&93.8 & 422.5 & 93.7 & 232.0   & 93.2 & 139.2\\
    \cline{2-9}\rule{0pt}{2.1ex}
    &\multirow{4}{*}{TF}&$\mathcal{I}_{\mbox{\tiny AIPW2}}$&93.1 & 412.9 & 92.7 & 230.5 & 92.6 & 144.1 \\
    & &$\mathcal{I}_{\mbox{\tiny AIPW2B}}$&93.4 & 418.7 & 93.1 & 235.5 & 93.2 & 149.4 \\
    & &$\mathcal{I}_{\mbox{\tiny MCP}}$&92.8 & 413.5 & 92.6 & 230.9 & 92.4 & 144.5 \\
    & &$\mathcal{I}_{\mbox{\tiny MCPB}}$&93.6 & 423.7 & 93.5 & 238.2 & 93.3 & 151.2 \\
    \cline{2-9}\rule{0pt}{2.1ex}
    &\multirow{6}{*}{FT}&$\mathcal{I}_{\mbox{\tiny IPW2}}$&81.5 & 407.5 & 54.9 & 235.3 & 23.9 & 157.8 \\
    & &$\mathcal{I}_{\mbox{\tiny PELR}}$&81.6 & 408.5 & 54.9 & 235.9 & 23.9 & 158.2 \\
    & &$\mathcal{I}_{\mbox{\tiny AIPW2}}$&93.1 & 410.3 & 93.4 & 225.2 & 93.1 & 135.2 \\
    & &$\mathcal{I}_{\mbox{\tiny AIPW2B}}$&93.5 & 413.7 & 93.5 & 227.0   & 93.2 & 136.3 \\
    & &$\mathcal{I}_{\mbox{\tiny MCP}}$&92.4 & 404.5 & 92.5 & 222.4 & 92.5 & 133.8\\
    & &$\mathcal{I}_{\mbox{\tiny MCPB}}$&93.9 & 422.7 & 93.7 & 231.9 & 93.3 & 139.2\\
    \hline
    \label{CIt05}
    \end{tabular}
\end{table}

\begin{table}[pt]
    \caption{$\%CP$ and $AL(\times100)$ for $95\%$ Confidence Intervals when $t=0.7$}
    \centering
    \begin{tabular}{cclrrrrrr}
    \hline\rule{0pt}{2.1ex}
    \multirow{2}{*}{n}& \multirow{2}{*}{Scenario}&\multirow{2}{*}{Estimator}&\multicolumn{2}{c}{$\rho=0.3$}&\multicolumn{2}{c}{$\rho=0.5$}&\multicolumn{2}{c}{$\rho=0.7$}\\
    \cline{4-9}\rule{0pt}{2.1ex}
    & & &$\% CP$ & $AL$ &$\% CP$ & $AL$& $\% CP$ & $AL$\\
    \hline\rule{0pt}{2.4ex}
    \multirow{16}{*}{100}&\multirow{6}{*}{TT}&$\mathcal{I}_{\mbox{\tiny IPW2}}$&93.2 & 852.5 & 93.7 & 474.9 & 93.4 & 294.8 \\
    & &$\mathcal{I}_{\mbox{\tiny PELR}}$&93.6 & 860.7 & 93.9 & 479.1 & 93.4 & 297.2 \\
    & &$\mathcal{I}_{\mbox{\tiny AIPW2}}$&92.9 & 847.8 & 92.8 & 465.1 & 93.3 & 279.0   \\
    & &$\mathcal{I}_{\mbox{\tiny AIPW2B}}$&94.8 & 917.3 & 94.9 & 502.8 & 94.9 & 300.9 \\
    & &$\mathcal{I}_{\mbox{\tiny MCP}}$&92.8 & 864.9 & 93.6 & 485.8 & 94.7 & 307.1 \\
    & &$\mathcal{I}_{\mbox{\tiny MCPB}}$&96.1 & 979.7 & 96.1 & 536.7 & 96.1 & 321.0  \\
    \cline{2-9}\rule{0pt}{2.1ex}
    &\multirow{4}{*}{TF}&$\mathcal{I}_{\mbox{\tiny AIPW2}}$&93.0   & 849.1 & 93.6 & 471.0   & 93.6 & 289.7 \\
    & &$\mathcal{I}_{\mbox{\tiny AIPW2B}}$&94.8 & 913.1 & 95.1 & 510.4 & 95.5 & 319.6 \\
    & &$\mathcal{I}_{\mbox{\tiny MCP}}$&93.2 & 858.7 & 93.6 & 478.4 & 93.7 & 296.8 \\
    & &$\mathcal{I}_{\mbox{\tiny MCPB}}$&96.2 & 959.4 & 95.4 & 535.8 & 96.2 & 335.9 \\
    \cline{2-9}\rule{0pt}{2.1ex}
    &\multirow{6}{*}{FT}&$\mathcal{I}_{\mbox{\tiny IPW2}}$&90.1 & 851.4 & 83.0   & 498.7 & 71.9 & 342.7 \\
    & &$\mathcal{I}_{\mbox{\tiny PELR}}$&90.3 & 859.5 & 83.6 & 503.5 & 72.5 & 346.3 \\
    & &$\mathcal{I}_{\mbox{\tiny AIPW2}}$&93.7 & 851.5 & 93.7 & 467.1 & 93.9 & 280.2 \\
    & &$\mathcal{I}_{\mbox{\tiny AIPW2B}}$&94.7 & 907.6 & 95.0   & 497.5 & 95.3 & 297.8 \\
    & &$\mathcal{I}_{\mbox{\tiny MCP}}$&93.2 & 838.1 & 93.5 & 464.7 & 94.3 & 285.6 \\
    & &$\mathcal{I}_{\mbox{\tiny MCPB}}$&95.8 & 977.9 & 96.0   & 535.8 & 96.2 & 320.5\\
    \hline\rule{0pt}{2.1ex}
    \multirow{16}{*}{200}&\multirow{6}{*}{TT}&$\mathcal{I}_{\mbox{\tiny IPW2}}$&94.0   & 610.9 & 95.4 & 338.3 & 94.7 & 207.5 \\
    & &$\mathcal{I}_{\mbox{\tiny PELR}}$&94.2 & 614.1 & 95.4 & 339.9 & 94.6 & 208.5 \\
    & &$\mathcal{I}_{\mbox{\tiny AIPW2}}$&93.7 & 608.7 & 94.1 & 333.8 & 94.2 & 200.2 \\
    & &$\mathcal{I}_{\mbox{\tiny AIPW2B}}$&94.7 & 626.6 & 94.9 & 343.5 & 95.0   & 205.8 \\
    & &$\mathcal{I}_{\mbox{\tiny MCP}}$&93.7 & 613.3 & 94.5 & 339.8 & 95.1 & 208.7 \\
    & &$\mathcal{I}_{\mbox{\tiny MCPB}}$&95.3 & 645.3 & 95.3 & 353.8 & 95.6 & 212.0 \\
    \cline{2-9}\rule{0pt}{2.1ex}
    &\multirow{4}{*}{TF}&$\mathcal{I}_{\mbox{\tiny AIPW2}}$&93.8 & 609.4 & 94.9 & 336.4 & 94.8 & 204.9 \\
    & &$\mathcal{I}_{\mbox{\tiny AIPW2B}}$&95.2 & 627.5 & 95.6 & 348.4 & 95.5 & 214.9 \\
    & &$\mathcal{I}_{\mbox{\tiny MCP}}$&94.5 & 612.7 & 95.0   & 338.6 & 94.8 & 207.0   \\
    & &$\mathcal{I}_{\mbox{\tiny MCPB}}$&95.5 & 642.5 & 96.1 & 356.3 & 96.1 & 219.7 \\
    \cline{2-9}\rule{0pt}{2.1ex}
    &\multirow{6}{*}{FT}&$\mathcal{I}_{\mbox{\tiny IPW2}}$&88.7 & 608.9 & 75.3 & 356.7 & 52.9 & 245.1 \\
    & &$\mathcal{I}_{\mbox{\tiny PELR}}$&88.9 & 612.0   & 75.4 & 358.6 & 53.3 & 246.6 \\
    & &$\mathcal{I}_{\mbox{\tiny AIPW2}}$&94.3 & 608.0   & 94.6 & 333.4 & 94.9 & 199.9 \\
    & &$\mathcal{I}_{\mbox{\tiny AIPW2B}}$&95.3 & 622.1 & 95.1 & 341.1 & 95.4 & 204.4 \\
    & &$\mathcal{I}_{\mbox{\tiny MCP}}$&94.0   & 599.7 & 94.3 & 331.3 & 95.2 & 201.7 \\
    & &$\mathcal{I}_{\mbox{\tiny MCPB}}$&95.4 & 644.3 & 95.6 & 353.1 & 96.0   & 211.5\\
    \hline\rule{0pt}{2.1ex}
    \multirow{16}{*}{400}&\multirow{6}{*}{TT}&$\mathcal{I}_{\mbox{\tiny IPW2}}$&94.5 & 434.7 & 94.8 & 241.0   & 94.4 & 148.0   \\
    & &$\mathcal{I}_{\mbox{\tiny PELR}}$&94.5 & 435.9 & 94.8 & 241.6 & 94.4 & 148.4 \\
    & &$\mathcal{I}_{\mbox{\tiny AIPW2}}$&94.2 & 432.3 & 94.4 & 237.1 & 94.1 & 142.1 \\
    & &$\mathcal{I}_{\mbox{\tiny AIPW2B}}$&94.7 & 437.6 & 94.6 & 239.9 & 94.5 & 143.8 \\
    & &$\mathcal{I}_{\mbox{\tiny MCP}}$&94.4 & 436.2 & 94.7 & 241.8 & 94.6 & 148.1 \\
    & &$\mathcal{I}_{\mbox{\tiny MCPB}}$&95.0   & 444.7 & 95.1 & 243.9 & 95.0   & 146.2\\
    \cline{2-9}\rule{0pt}{2.1ex}
    &\multirow{4}{*}{TF}&$\mathcal{I}_{\mbox{\tiny AIPW2}}$&94.4 & 434.2 & 94.5 & 240.3 & 94.4 & 146.9 \\
    & &$\mathcal{I}_{\mbox{\tiny AIPW2B}}$&94.9 & 440.1 & 94.9 & 244.7 & 94.6 & 151.0   \\
    & &$\mathcal{I}_{\mbox{\tiny MCP}}$&94.5 & 435.6 & 94.7 & 241.2 & 94.3 & 147.7 \\
    & &$\mathcal{I}_{\mbox{\tiny MCPB}}$&95.0   & 446.2 & 95.2 & 247.7 & 95.6 & 152.8 \\
    \cline{2-9}\rule{0pt}{2.1ex}
    &\multirow{6}{*}{FT}&$\mathcal{I}_{\mbox{\tiny IPW2}}$&82.8 & 431.1 & 55.9 & 252.4 & 24.7 & 173.4 \\
    & &$\mathcal{I}_{\mbox{\tiny PELR}}$&82.9 & 432.3 & 56.0   & 253.1 & 25.2 & 173.9 \\
    & &$\mathcal{I}_{\mbox{\tiny AIPW2}}$&94.6 & 430.9 & 94.7 & 236.3 & 94.5 & 141.7 \\
    & &$\mathcal{I}_{\mbox{\tiny AIPW2B}}$&94.6 & 434.7 & 94.7 & 238.3 & 94.9 & 142.8 \\
    & &$\mathcal{I}_{\mbox{\tiny MCP}}$&94.2 & 425.7 & 94.5 & 234.7 & 94.7 & 142.3\\
    & &$\mathcal{I}_{\mbox{\tiny MCPB}}$&94.9 & 444.3 & 94.9 & 243.3 & 95.3 & 145.8\\
    \hline
    \label{CIt07}
    \end{tabular}
\end{table}

\subsection{Hypothesis tests}

The basic hypothesis test problem in causal inference is to test whether there is a treatment effect, i.e., to test $H_0: \theta=0$ versus $H_1: \theta \ne 0$. The power of a test (at the $0.05$ level) can be computed through the rejection rate of the corresponding $95\%$ confidence interval not containing zero over repeated samples. Specifically, for a test method corresponding to a confidence interval $\mathcal{I}$, we have
 \[
    \text{Rejection Rate}=\frac{1}{n_{sim}}\sum_{s=1}^{n_{sim}}I\left(0\notin \mathcal{I}^{(s)}\right)\, .
 \]
We consider four test methods based on the confidence intervals $\mathcal{I}_{\mbox{\tiny AIPW2}}$, $\mathcal{I}_{\mbox{\tiny AIPW2B}}$, $\mathcal{I}_{\mbox{\tiny MCP}}$ and $\mathcal{I}_{\mbox{\tiny MCPB}}$, respectively.
When the true value of $\theta$ equals zero, the rejection rate is an approximation to the type I error probability; otherwise, the rejection rate is an approximation to the power of the test for a given value of the true ATE, $\theta^0$. 

The sample data are generated following the same setup in Section \ref{simCP}, except that the two outcome regression models are modified as 
\[
m_1:\;\;\;Y_{1j}=\theta^0+4.5+x_{j1}-2x_{j2}+3x_{j3}+a_1\epsilon_j\]  and
\[m_0:\;\;\;Y_{0j}=3.88+x_{j1}+x_{j2}+2x_{j3}+a_0\epsilon_j\, ,
\]
where $\theta^0$ is the assigned value of the true ATE. The setting allows $\theta^0$ to vary from $0$ to $3$ to show the pattern of the power function of the test. 

\begin{figure}
\begin{tabular}{ccc}
  \includegraphics[width=0.33\textwidth]{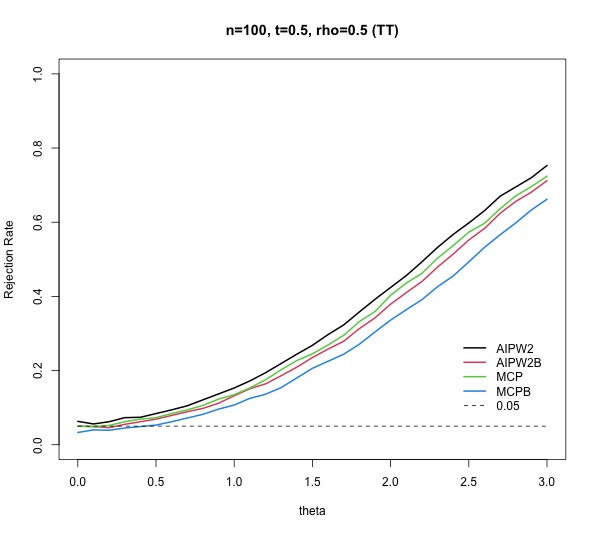} & \includegraphics[width=0.33\textwidth]{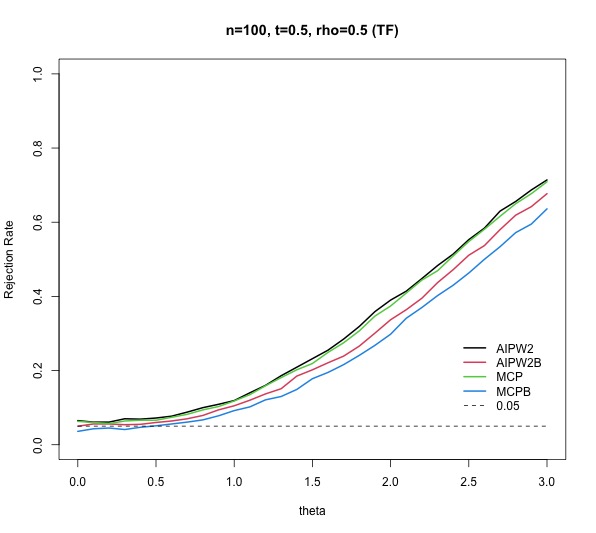}&
  \includegraphics[width=0.33\textwidth]{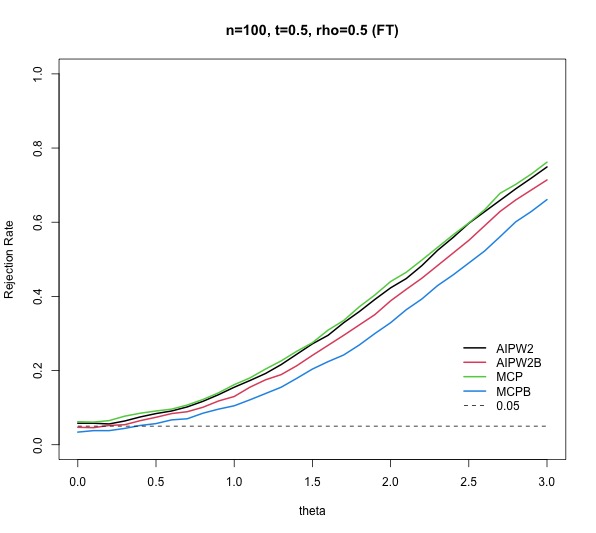}\\
  \includegraphics[width=0.33\textwidth]{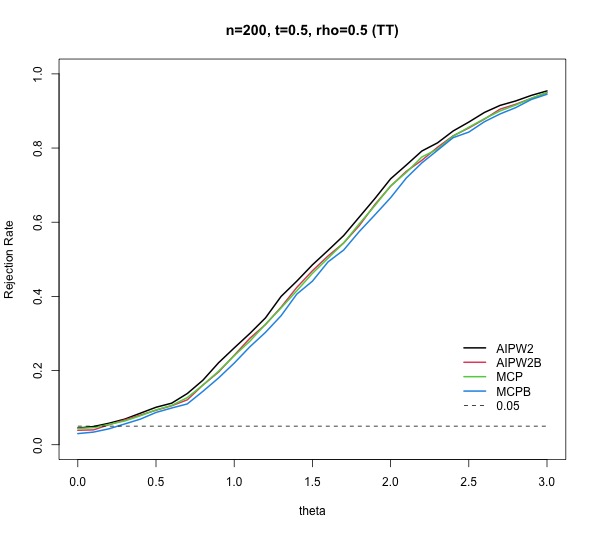} & \includegraphics[width=0.33\textwidth]{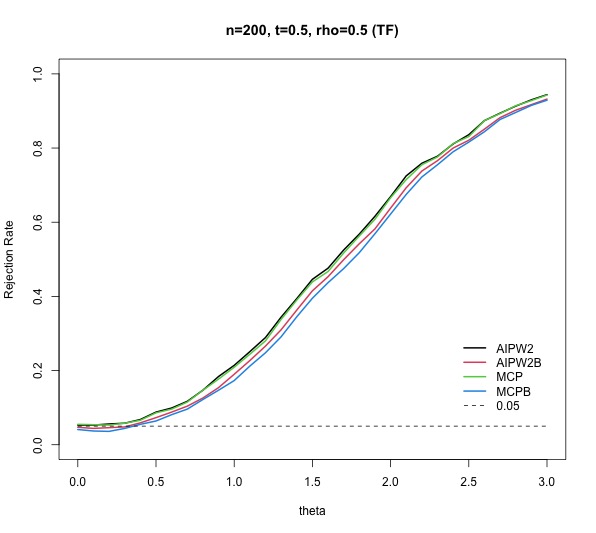}&
  \includegraphics[width=0.33\textwidth]{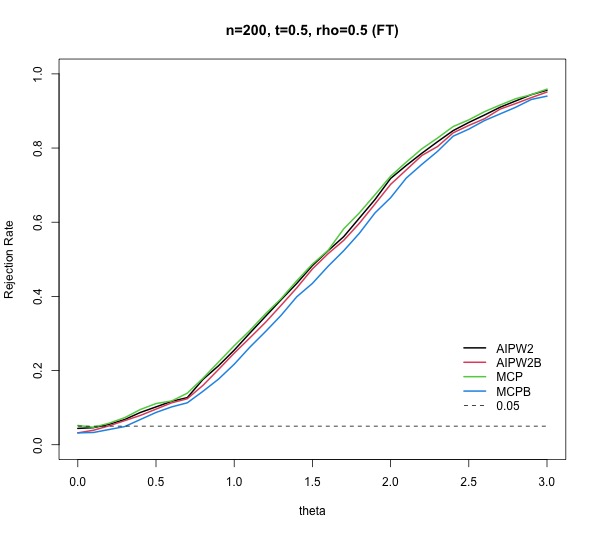}\\
  \includegraphics[width=0.33\textwidth]{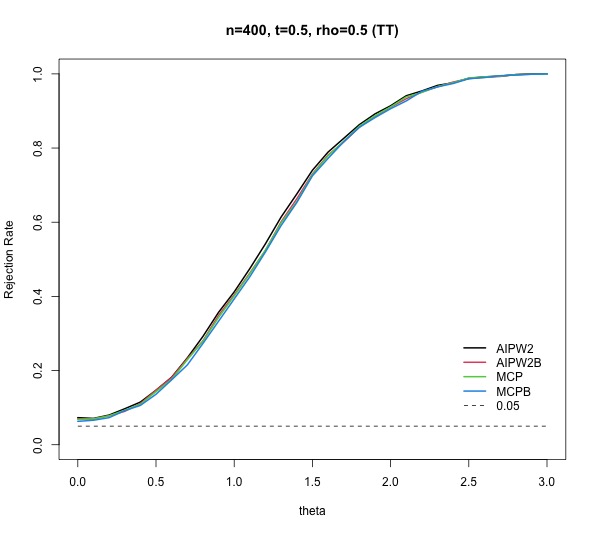} & \includegraphics[width=0.33\textwidth]{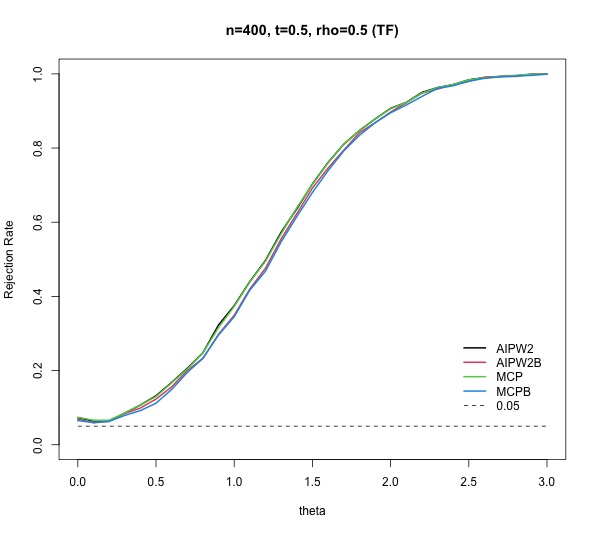}&
  \includegraphics[width=0.33\textwidth]{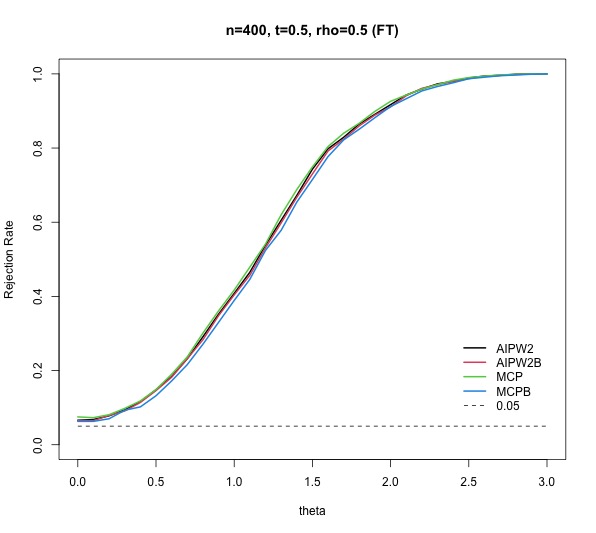}
\end{tabular}
\caption{Power functions of tests when $t=0.5$ and $\rho=0.5$}
\label{rejectiont05rho05}
\end{figure}

\begin{figure}
    \centering
    \includegraphics[width=0.8\textwidth]{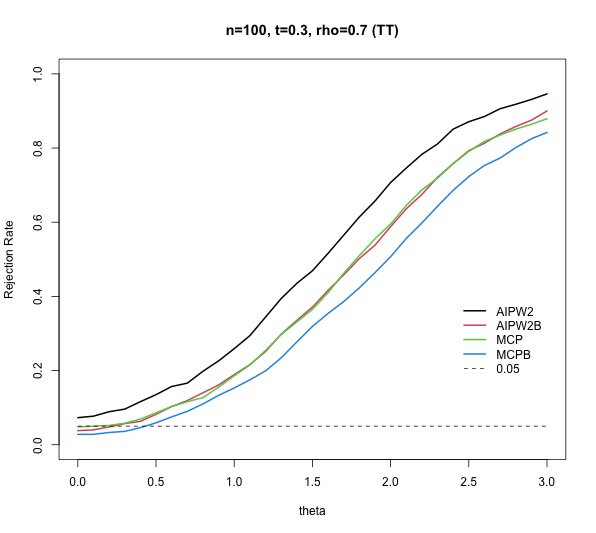}
    \caption{Power functions of tests with $(n,t,\rho)=(100, 0.3, 0.7)$ and ``TT''}
    \label{rejectionrho07}
\end{figure}

Figure \ref{rejectiont05rho05} depicts the power functions of tests for different sample sizes and model specifications with $t=0.5$ and $\rho=0.5$. 
For $n=100$, it is clear that the two curves for the bootstrap-calibrated confidence intervals are below the other two, implying that they have smaller type I error probabilities and test powers, though there is not much difference between all four curves. 
If the sample size increases, the four curves get closer and become indistinguishable when $n=400$.
The test powers of the four methods all go up as $\theta^0$ departs from $0$ for all the cases, meaning that they are working well; especially when $n=400$, the test powers increase to $1$ rapidly. 
The test using $\mathcal{I}_{\mbox{\tiny AIPW2}}$ seems to have larger powers than others in many cases. However, a further examination reveals that it also has inflated type I error probabilities in some cases, as shown in Figure \ref{rejectionrho07}. 

\section{Additional Remarks}
\label{section6}

Estimation methods discussed in this paper rely on the validity of the assumptions {\bf A1} and {\bf A2}. The most essential one, the SITA assumption {\bf A1}, however, is not testable with the given sample data. 
A general guideline for the practical use of the estimation methods is to include potential predictors for both the treatment assignment mechanism and the outcome variables during the data collection process. 
Moreover, \cite{Brookhart2006} demonstrated that incorporating outcome-related variables into the propensity score model can improve the efficiency of estimation while avoiding additional biases. 
The failure of the positivity assumption {\bf A2} means certain subjects will have no chance to be included in the treatment group, a scenario similar to the under-coverage problem in survey sampling where some subjects have no chance to be included in the sample. Under-coverage problems are seemingly simple but notoriously difficult to address in survey sampling. They require additional information on the unknown population.

The simulation results seem to suggest that none of the methods uniformly outperforms the others for all scenarios. 
Our proposed PEL methods, however, have two major advantages. The maximum PEL estimators are obtained through a constrained maximization procedure, which allows the use of any suitable auxiliary information through the inclusion of additional constraints. One obvious direction of extending the methods of this paper is to construct multiply robust estimators of the ATE by incorporating multiple working models \citep{Han2013}. The PEL ratio confidence intervals are range-respecting and transformation invariant, which is an attractive feature for scenarios where, for instance, the true value of the ATE is confined within a particular range such as $(-1,1)$. PEL ratio confidence intervals under the multiple robustness framework is a topic of future investigation.

\section{Appendix}
\label{Sec.appendix}

\subsection{Regularity conditions}
\label{RC}

Let $m_1(\boldsymbol{x}, \boldsymbol{\beta}_1)$ and $m_0(\boldsymbol{x}, \boldsymbol{\beta}_0)$ be respectively the mean functions of the outcome regression models for $Y_1$ and $Y_0$ given the covariate vector $\boldsymbol{x}$.
The probability limit of the estimator $\hat{\boldsymbol{\beta}}_i$ for the vector of model parameters $\boldsymbol{\beta}_i$ under the assumed working model using the observed sample data is denoted as $\boldsymbol{\beta}_i^*$ for $i=1,0$.

\begin{itemize}
    \item [{\bf R1}] The treatment indicator $T$ satisfies $E(T)=t\in (0,1)$.

    \item [{\bf R2}] The population satisfies $\operatorname{E}({Y}_{1}^2)<\infty$, $\operatorname{E}({Y}_{0}^2)<\infty$, and $\operatorname{E}(||\boldsymbol{x}||^2)<\infty$, where $||\cdot||$ denotes the $l_2$-norm.

    \item [{\bf R3}] The population and the mean functions satisfy $\operatorname{E}\{m_1^2(\boldsymbol{x},\boldsymbol{\beta}_1^*)\}<\infty$ and $\operatorname{E}\{m_0^2(\boldsymbol{x},\boldsymbol{\beta}_0^*)\}<\infty$.

    \item [{\bf R4}] For each $\boldsymbol{x}$, $\partial m_i(\boldsymbol{x}, \boldsymbol{\beta}_i)/\partial\boldsymbol{\beta}_i$ is continuous in $\boldsymbol{\beta}_i$ and $||\partial m_i(\boldsymbol{x}, \boldsymbol{\beta}_i)/\partial\boldsymbol{\beta}_i||\leq h_i(\boldsymbol{x}, \boldsymbol{\beta}_i)$ for $\boldsymbol{\beta}_i$ in the neighborhood of $\boldsymbol{\beta}_i^*$, and $\operatorname{E}\{h_i(\boldsymbol{x}, \boldsymbol{\beta}_i^*)\}<\infty$, for $i=0,1$.

     \item [{\bf R5}] For each $\boldsymbol{x}$, $\partial^2 m_i(\boldsymbol{x}, \boldsymbol{\beta}_i)/\partial\boldsymbol{\beta}_i\partial\boldsymbol{\beta}_i^\top$ is continuous in $\boldsymbol{\beta}_i$ and $\operatorname{max}_{j,l}|\partial^2 m_i(\boldsymbol{x}, $\\$\boldsymbol{\beta}_i)/\partial\boldsymbol{\beta}_{ij}\partial\boldsymbol{\beta}_{il}|\leq k_i(\boldsymbol{x}, \boldsymbol{\beta}_i)$ for $\boldsymbol{\beta}_i$ in the neighborhood of $\boldsymbol{\beta}_i^*$, and $\operatorname{E}\{k_i(\boldsymbol{x}, $\\$\boldsymbol{\beta}_i^*)\}<\infty$, for $i=0,1$, where $|\cdot|$ denotes the absolute value.

     \item[{\bf R6}] There exist $c_1$ and $c_2$ such that $0<c_1\leq \tau_j^0\leq c_2<1$ for all units $j$, where $\tau_j^0=\tau(\boldsymbol{x}_j;\boldsymbol{\alpha}^0)$ is the propensity score under the true propensity score model and $\boldsymbol{\alpha}^0$ is the true value for the vector of propensity score model parameters $\boldsymbol{\alpha}$.
\end{itemize}

Condition {\bf R1} is commonly used in practice. 
Note that the sample size of the treatment group $n_1 = \sbsj \Ri$ and then, $n_1/n=\sbsj \Ri/n$, where $n$ is the sample size. 
Under condition {\bf R1}, we have $n_1/n$ converges to $t\in(0,1)$. 
Thus, there are no essential differences among $O_p(n_1^{-1/2})$, $O_p(n_0^{-1/2})$ and $O_p(n^{-1/2})$.
Conditions {\bf R2} and {\bf R3} are the typical finite moment conditions. If the outcome regression models are linear, then 
$\operatorname{E}(||\boldsymbol{x}||^2)<\infty$ from {\bf R2} implies {\bf R3}. 
As stated in Lemma 11.2 of \cite{owen2001}, if $Y_1, \dots, Y_n$ are independent and identically distributed as the random variable $Y$, and $\operatorname{E}(Y^2)<\infty$, then $\operatorname{max}_{1\leq i\leq n}|Y_i|=o_p(\sqrt{n})$.
Conditions {\bf R2} and {\bf R3} play an important role in establishing the $O_p(n^{-1/2})$ order of the Lagrange multiplier for the constrained maximization problems. 
Conditions {\bf R4} and {\bf R5} are the usual smoothness and boundedness conditions \citep{Wu2001, Chen2020}. They are automatically satisfied under linear outcome regression models.
Condition {\bf R6} states that each unit in the initial sample has a non-trivia probability of being assigned to either the treatment group or the control group.

\subsection{Sandwich variance estimators for IPW estimators}
\label{sandforIPW}
Let $\bPsi=(\mu_1, \theta, \balpha^\top)^\top $. 
The combination of estimators $\hat{\bPsi}=(\hat{\mu}_1, \hat{\theta}, \hbalpha^\top)^\top$, where $\hbalpha$ is the vector of estimators for the coefficients in the logistic regression model for propensity scores, is the solution to the estimating equations system 
\begin{equation}
    \boldsymbol{\mathrm{U}}(\bPsi)=\frac{1}{n}\sum_{j=1}^{n}\left(\begin{array}{c} \frac{\Ri\left(\Yti-\mu_1\right)}{\tau_{j}}+\Delta_1 \frac{\Ri-\tau_{j}}{\tau_{j}} \\
     \frac{(1-\Ri)\left(\Yci-(\mu_1-\theta)\right)}{1-\tau_{j}}+\Delta_0 \frac{\tau_{j}-\Ri}{1-\tau_{j}} \\
     \bz(\Ri-\tau_j)\end{array}\right)=\frac{1}{n}\sbsj\boldsymbol{\mathrm{U}}_j(\bPsi)=\boldsymbol{0}\, ,
    \notag
\end{equation}
where $\tau_j=\tau(\bx; \balpha)=\{1+\exp({\bz^\top\balpha})\}^{-1}\exp({\bz^\top\balpha})$, as specified by logistic regression models. 
The H\'{a}jek-type IPW estimators $\hat{\mu}_1=\mutH$ and $\hat{\theta}=\thetaH$ are obtained when setting $\boldsymbol{\Delta}=(\Delta_1, \Delta_0)=(0,0)$ in the estimating equations system. 
Similarly, the IPW estimators $\hat{\mu}_1=\mutIPW$ and $\hat{\theta}=\thetaIPW$ can be derived by choosing $\boldsymbol{\Delta}=(\Delta_1, \Delta_0)=(\mu_1,\mu_1-\theta)$. 
This formulation is similar to the one used for proving Theorem 1 in \cite{Chen2020}. 
Let $\bPsi^{0}=(\mu_1^0, \theta^0, \{\balpha^{0}\}^\top)^\top $ denote the vector of true values. 
Under the SITA assumption {\bf A1} and the assumed propensity score model, we have $\operatorname{E}(\boldsymbol{\mathrm{U}}(\bPsi^0))=\boldsymbol{0}$ when $\boldsymbol{\Delta}=(0,0)$ or $\boldsymbol{\Delta}=(\mu_1,\mu_1-\theta)$.
It follows that the estimator $\hat{\bPsi}$ is an $m$-estimator.
Hence, $\hat{\bPsi} \stackrel{P}{\rightarrow} \bPsi^{0}$ (component-wise) under the regularity conditions {\bf R2} and {\bf R6}.
Therefore, both types of IPW estimators are valid when the propensity score model is correctly specified.
We have $\boldsymbol{\mathrm{U}}(\hat{\bPsi})=0$ and $\hat{\bPsi} = \bPsi^{0}+o_p(1)$ (component-wise).

Under the regularity conditions {\bf R2} and {\bf R6} and the fact that $\operatorname{E}(\boldsymbol{\mathrm{U}}(\bPsi^0))=\boldsymbol{0}$, we have $\boldsymbol{\mathrm{U}}(\boldsymbol{\Psi}^0)=O_p(n^{-1/2})$. 
Taking the Taylor expansion to $\boldsymbol{\mathrm{U}}(\hat{\bPsi})$ at $\bPsi=\bPsi^{0}$ yields
\[
\scalemath{0.86}{
    \hat{\bPsi}-\bPsi^{0}=-\left[\boldsymbol{\mathrm{H}}\left(\bPsi^{0}\right)\right]^{-1} \boldsymbol{\mathrm{U}}\left(\bPsi^{0}\right)+o_{p}\left(n^{-1 / 2}\right)=-\left[\operatorname{E}\left\{\boldsymbol{\mathrm{H}}\left(\bPsi^{0}\right)\right\}\right]^{-1} \boldsymbol{\mathrm{U}}\left(\bPsi^{0}\right)+o_{p}\left(n^{-1 / 2}\right)}\, ,
\]
where $\boldsymbol{\mathrm{H}}(\bPsi)=\partial \boldsymbol{\mathrm{U}}(\bPsi)/\partial \bPsi$.
It follows that $\hat{\bPsi}-\bPsi^{0}=O_p(n^{-1/2})$.
Also, the theoretical asymptotic variance of $\hat{\bPsi}$ takes the sandwich form of
\[
    \operatorname{Var}\left(\hat{\bPsi}\right)=\left[\operatorname{E}\left\{\boldsymbol{\mathrm{H}}\left(\bPsi^{0}\right)\right\}\right]^{-1} \operatorname{Var}\left\{\boldsymbol{\mathrm{U}}\left(\bPsi^{0}\right)\right\}\left[\operatorname{E}\left\{\boldsymbol{\mathrm{H}}\left(\bPsi^{0}\right)\right\}^\top\right]^{-1}\, ,
\]
where $\operatorname{Var}\{\boldsymbol{\mathrm{U}}(\bPsi^{0})\}=n^{-1}\operatorname{E}\{\boldsymbol{\mathrm{U}}_j(\bPsi^{0})\boldsymbol{\mathrm{U}}_j^\top(\bPsi^{0})\}$ and $\boldsymbol{\mathrm{U}}_j(\bPsi)$ is defined as part of $\boldsymbol{\mathrm{U}}(\bPsi)$ at the beginning of this section. This implies that the variances of the IPW estimators $\thetaIPW$ and $\thetaH$ can be estimated by the second diagonal element of the corresponding sandwich variance estimator given by 
\[
    \operatorname{var}\left(\hat{\bPsi}\right)=\left\{\boldsymbol{\mathrm{H}}\left(\hat{\bPsi}\right)\right\}^{-1} \left\{\frac{1}{n^2}\sum_{j=1}^n\boldsymbol{\mathrm{U}}_j\left(\hat{\bPsi}\right)\boldsymbol{\mathrm{U}}_j^\top\left(\hat{\bPsi}\right)\right\}\left\{\boldsymbol{\mathrm{H}}^\top\left(\hat{\bPsi}\right)\right\}^{-1}\, ,
\]
where $\operatorname{var}(\cdot)$ denotes a variance estimator.

\subsection{Proof of Proposition \ref{corollary1}}
\label{proofAIPW}

\begin{proof}
Let $\boldsymbol{\Psi}=(\mu_1, \theta, \Bar{m}_1, \Bar{m}_0, \balpha^\top, \boldsymbol{\beta}_1^\top, \boldsymbol{\beta}_0^\top)^\top$, where $\Bar{m}_1$ and $\Bar{m}_0$ are two additional nuisance parameters. 
The estimator $\hat{\boldsymbol{\Psi}}=(\hat{\mu}_1, \hat{\theta}, \Bar{\hat{m}}_1, \Bar{\hat{m}}_0, \hbalpha^\top, \hat{\boldsymbol{\beta}}_1^\top, \hat{\boldsymbol{\beta}}_0^\top)^\top$ is the solution to the set estimating equations
\begin{equation}
\scalemath{0.88}{
    \boldsymbol{\mathrm{U}}(\boldsymbol{\Psi})=\frac{1}{n}\sum_{j=1}^{n}\left(\begin{array}{c} \frac{\Ri}{\tau_{j}}\left(\Yti-m_{1j}+\Bar{m}_1-\mu_1\right)+\Delta_1 \frac{\Ri-\tau_{j}}{\tau_{j}} \\
     \frac{(1-\Ri)}{1-\tau_{j}}\left(\Yci-m_{0j}+\Bar{m}_0-(\mu_1-\theta)\right)+\Delta_0 \frac{\tau_{j}-\Ri}{1-\tau_{j}} \\
     m_{1j}-\Bar{m}_1\\
     m_{0j}-\Bar{m}_0\\
     \bz(\Ri-\tau_j)\\
     \Ri(\Yti-m_{1j})\bz\\
     (1-\Ri)(\Yci-m_{0j})\bz\end{array}\right)=\frac{1}{n}\sbsj\boldsymbol{\mathrm{U}}_j(\boldsymbol{\Psi})=\boldsymbol{0}}\, .
    \label{Uequation}
\end{equation}
Without loss of generality, we assume that the outcome regression models are linear. The arguments work for nonlinear outcome regression models.

The estimators $\hat{\mu}_1=\mutDRT$ and $\hat{\theta}=\thetaDRT$ are obtained by setting $\boldsymbol{\Delta}=(\Delta_1, \Delta_0)=(0,0)$. 
Similarly, the estimators $\hat{\mu}_1=\mutDRO$ and $\hat{\theta}=\thetaDRO$ can be derived by choosing $\boldsymbol{\Delta}=(\Delta_1, \Delta_0)=(\mu_1-\Bar{m}_1,\mu_1-\theta-\Bar{m}_0)$. 
Again, the formulation of (\ref{Uequation}) is similar to the one used for proving Theorem 1 in \cite{Chen2020}. 

Under the current framework, we have $\hat{\boldsymbol{\alpha}}-\boldsymbol{\alpha}^*=O_p(n^{-1/2})$ for some value $\boldsymbol{\alpha}^*$ regardless of the correct specification of the propensity score model \citep{White1982}.
Similarly, it is assumed that the estimator $\hat{\boldsymbol{\beta}}_i$ for $\boldsymbol{\beta}_i$ satisfies $\hat{\boldsymbol{\beta}}_i-\boldsymbol{\beta}_i^*=O_p(n^{-1/2})$ for some value $\boldsymbol{\beta}_i^*$ regardless of the true outcome regression model, for $i=0, 1$. 
Let $\mu_1^0$ and $\theta^0$ denote the true values of $\mu_1$ and $\theta$. Let $\boldsymbol{\Psi}^{0}=(\mu_1^0, \theta^0, \operatorname{E}(m_{1j}^*), \operatorname{E}(m_{0j}^*), {\boldsymbol{\alpha}^*}^\top, {\boldsymbol{\beta}_1^*}^\top, {\boldsymbol{\beta}_0^*}^\top)^\top$.
We have $\operatorname{E}(\boldsymbol{\mathrm{U}}(\boldsymbol{\Psi}^0))=\boldsymbol{0}$ if either of the propensity score model and the set of outcome regression models is correctly specified. 
The estimator $\hat{\boldsymbol{\Psi}}$ is an $m$-estimator \citep{tsiatis2006semi} and $\thetaDRO$ and $\thetaDRT$ are doubly robust.

It follows that we have $\boldsymbol{\mathrm{U}}(\hat{\boldsymbol{\Psi}})=0$ and $\hat{\boldsymbol{\Psi}} = \boldsymbol{\Psi}^{0}+o_p(1)$ (component-wise) under the regularity conditions {\bf R2}-{\bf R3} and {\bf R6} when one of the two sets of models is correctly specified.

By the central limit theorem and under regularity conditions {\bf R2}-{\bf R3} and {\bf R6}, the result $\operatorname{E}(\boldsymbol{\mathrm{U}}(\boldsymbol{\Psi}^0))=\boldsymbol{0}$ leads to $\boldsymbol{\mathrm{U}}(\boldsymbol{\Psi}^0)=O_p(n^{-1/2})$. 
Given that one of the two sets of models is correctly specified, taking the Taylor  expansion of $\boldsymbol{\mathrm{U}}(\hat{\boldsymbol{\Psi}})$ at $\bPsi=\boldsymbol{\Psi}^{0}$ yields
\[\scalemath{0.86}{
    \hat{\boldsymbol{\Psi}}-\boldsymbol{\Psi}^{0}=-\left[\boldsymbol{\mathrm{H}}\left(\boldsymbol{\Psi}^{0}\right)\right]^{-1} \boldsymbol{\mathrm{U}}\left(\boldsymbol{\Psi}^{0}\right)+o_{p}\left(n^{-1 / 2}\right)=-\left[\operatorname{E}\left\{\boldsymbol{\mathrm{H}}\left(\boldsymbol{\Psi}^{0}\right)\right\}\right]^{-1} \boldsymbol{\mathrm{U}}\left(\boldsymbol{\Psi}^{0}\right)+o_{p}\left(n^{-1 / 2}\right)}\, ,
\]
where $\boldsymbol{\mathrm{H}}(\boldsymbol{\Psi})=\partial \boldsymbol{\mathrm{U}}(\boldsymbol{\Psi})/\partial \boldsymbol{\Psi}$.
The expansion leads to $\hat{\boldsymbol{\Psi}}-\boldsymbol{\Psi}^{0}=O_p(n^{-1/2})$.
The theoretical asymptotic variance of $\hat{\boldsymbol{\Psi}}$ takes the sandwich form of
\[
    \operatorname{Var}\left(\hat{\boldsymbol{\Psi}}\right)=\left[\operatorname{E}\left\{\boldsymbol{\mathrm{H}}\left(\boldsymbol{\Psi}^{0}\right)\right\}\right]^{-1} \operatorname{Var}\left\{\boldsymbol{\mathrm{U}}\left(\boldsymbol{\Psi}^{0}\right)\right\}\left[\operatorname{E}\left\{\boldsymbol{\mathrm{H}}\left(\boldsymbol{\Psi}^{0}\right)\right\}^\top\right]^{-1}\, ,
\]
where $\operatorname{Var}\{\boldsymbol{\mathrm{U}}(\boldsymbol{\Psi}^{0})\}=n^{-1}\operatorname{E}\{\boldsymbol{\mathrm{U}}_j(\boldsymbol{\Psi}^{0})\boldsymbol{\mathrm{U}}_j^\top(\boldsymbol{\Psi}^{0})\}$. Consequently,  the variances of the augmented IPW estimators $\thetaDRO$ and $\thetaDRT$ can be estimated by the second diagonal element of the corresponding sandwich variance estimator 
\[
    \operatorname{var}\left(\hat{\boldsymbol{\Psi}}\right)=\left\{\boldsymbol{\mathrm{H}}\left(\hat{\boldsymbol{\Psi}}\right)\right\}^{-1} \left\{\frac{1}{n^2}\sum_{j=1}^n\boldsymbol{\mathrm{U}}_j\left(\hat{\boldsymbol{\Psi}}\right)\boldsymbol{\mathrm{U}}_j^\top\left(\hat{\boldsymbol{\Psi}}\right)\right\}\left\{\boldsymbol{\mathrm{H}}^\top\left(\hat{\boldsymbol{\Psi}}\right)\right\}^{-1}\, ,
\]
where $\operatorname{var}(\cdot)$ denotes a variance estimator.
Note that the aforementioned variance estimator for the augmented IPW $\thetaDRO$ or $\thetaDRT$ is doubly robust.
As a matter of fact, the sandwich variance estimator is valid even if both sets of models are misspecified, since we can simply replace $(\mu_1^0, \theta^0)$ in the above proof with the limit $(\mu_1^*, \theta^*)$ under misspecified models. 
\end{proof}

\subsection{Proof of Theorem \ref{PELWOMC}}
\label{PELWOMCProof}

\begin{proof}
Under the regularity condition {\bf R1}, there is no need to distinguish among $O_p(n^{-1/2})$, $O_p(n_1^{-1/2})$, and $O_p(n_0^{-1/2})$. 
Following \cite{Wu2012}, the normalization constraint (\ref{jnorm}) and the parameter constraint (\ref{para}) are equivalent to
\begin{equation}
\begin{array}{l}\sum_{i=0}^{1} w_{i} \sum_{j=1}^{n_{i}} p_{i j}=1\,, \\ \sum_{i=0}^{1} w_{i} \sum_{j=1}^{n_{i}} p_{i j} \boldsymbol{u}_{i j}=\boldsymbol{0}\,,
\end{array}
\notag
\end{equation}
where $\boldsymbol{u}_{i j}=\boldsymbol{Z}_{ij}-\boldsymbol\eta$ with $\boldsymbol{Z}_{1j}=(1, Y_{1j}/w_1)^\top$, $\boldsymbol{Z}_{0j}=(0, -Y_{0j}/w_0)^\top$, and $\boldsymbol\eta=(w_1, \theta)^\top$. 

Recall that $\scalemath{0.99}{\hat{\mbox{\boldmath{$p$}}}_1(\theta) = (\hat{p}_{11}(\theta),\cdots,\hat{p}_{1n_1}(\theta))^\top}$ and $\scalemath{0.99}{\hat{\mbox{\boldmath{$p$}}}_0(\theta) = (\hat{p}_{01}(\theta),\cdots,\hat{p}_{0n_0}(\theta))^\top}$ denote the maximizer of the joint pseudo-empirical likelihood function ({\ref{jointPEL}}) subject to the normalization constraints (\ref{jnorm}) and the parameter constraint (\ref{para}) for a fixed value of $\theta$. 
Using the Lagrange multiplier method, we have
$\hat{p}_{ij}(\theta)=\tilde{a}_{ij}/(1+\hat{\boldsymbol{\lambda}}^\top\boldsymbol{u}_{ij})$ for $i=0,1$ and $\hat{\boldsymbol{\lambda}}$ is the solution to 
\begin{equation}
\sum_{i=0}^{1} w_{i} \sum_{j=1}^{n_{i}} \frac{\tilde{a}_{ij}\boldsymbol{u}_{ij}}{(1+\boldsymbol{\lambda}^\top\boldsymbol{u}_{ij})}=\boldsymbol{0}\; .   
\label{require}
\end{equation}
The above equations can be further rewritten as 
\[
\sum_{i=0}^{1} w_{i} \sum_{j=1}^{n_{i}} \tilde{a}_{ij}\boldsymbol{u}_{ij}=\sum_{i=0}^{1} w_{i} \sum_{j=1}^{n_{i}} \frac{\tilde{a}_{ij}\boldsymbol{u}_{ij}\boldsymbol{u}^\top_{ij}}{(1+\boldsymbol{\lambda}^\top\boldsymbol{u}_{ij})}\boldsymbol{\lambda}\; .
\]
Let $\boldsymbol{U}=\sum_{i=0}^{1} w_{i} \sum_{j=1}^{n_{i}} \tilde{a}_{ij}\boldsymbol{u}_{ij} = (0, \thetaH-\theta)^\top$, which is of the order $O_p(n^{-1/2})$ under the assumed propensity score model if $\theta=\theta^0+O_p(n^{-1/2})$. We have
\[
\left\|\sum_{i=0}^{1} w_{i} \sum_{j=1}^{n_{i}} \frac{\tilde{a}_{ij}\boldsymbol{u}_{ij}\boldsymbol{u}^\top_{ij}}{(1+\boldsymbol{\lambda}^\top\boldsymbol{u}_{ij})}\right\|\geq\frac{1}{1+\|\boldsymbol{\lambda}\|\max_{\{i, j\}}\|\boldsymbol{u}_{ij}\|}\left\|\sum_{i=0}^{1} w_{i} \sum_{j=1}^{n_{i}} \tilde{a}_{ij}\boldsymbol{u}_{ij}\boldsymbol{u}^\top_{ij}\right\|  \, ,  
\]
where $||\cdot||$ denotes the $\ell_2$-norm.
The above results imply that 
\[
    \|\boldsymbol{U}\|\geq \frac{1}{1+\|\boldsymbol{\lambda}\|\max_{\{i, j\}}\|\boldsymbol{u}_{ij}\|}\left\|\sum_{i=0}^{1} w_{i} \sum_{j=1}^{n_{i}} \tilde{a}_{ij}\boldsymbol{u}_{ij}\boldsymbol{u}^\top_{ij}\right\| \|\boldsymbol{\lambda}\|\; .
\]
It is known that $\|\boldsymbol{U}\|=O_p(n^{-1/2})$, and under the regularity condition {\bf R2}, $\max_{\{i, j\}}\|\boldsymbol{u}_{ij}\|=o_p(n^{1/2})$ and $\|\sum_{i=0}^{1} w_{i} \sum_{j=1}^{n_{i}} \tilde{a}_{ij}\boldsymbol{u}_{ij}\boldsymbol{u}^\top_{ij}\|=O_p(1)$. 
Therefore, we must have $\hat{\boldsymbol{\lambda}}=O_p(n^{-1/2})$. 
Since $\hat{\boldsymbol{\lambda}}^\top\boldsymbol{u}_{ij}=o_p(1)$ uniformly for all $i, j$, we have an expansion to $\hat{\boldsymbol{\lambda}}$ as 
\[
    \hat{\boldsymbol{\lambda}}=\left(\sum_{i=0}^{1} w_{i} \sum_{j=1}^{n_{i}} \tilde{a}_{ij}\boldsymbol{u}_{ij}\boldsymbol{u}^\top_{ij}\right)^{-1}\boldsymbol{U}+o_p(n^{-1/2})=\boldsymbol{D}^{-1}\boldsymbol{U}+o_p(n^{-1/2})\; ,
\]
where $\boldsymbol{D}=\sum_{i=0}^{1} w_{i} \sum_{j=1}^{n_{i}} \tilde{a}_{ij}\boldsymbol{u}_{ij}\boldsymbol{u}^\top_{ij}$ is a $2\times2$ matrix.
The expansion to  $\hat{\boldsymbol{\lambda}}$ is a crucial step to establish the asymptotic expansion to the PEL statistic $-2r_{\mbox{\tiny PEL}}(\theta)$, given below.
\begin{equation}
\begin{aligned}
-2r_{\mbox{\tiny PEL}}(\theta)&=2n \sum_{i=0}^{1} w_{i} \sum_{j=1}^{n_{i}} \tilde{a}_{ij}\log{(1+\hat{\boldsymbol{\lambda}}^\top\boldsymbol{u}_{ij}})\\
&=2n \sum_{i=0}^{1} w_{i} \sum_{j=1}^{n_{i}} \tilde{a}_{ij}\left(\hat{\boldsymbol{\lambda}}^\top\boldsymbol{u}_{ij}-\frac{1}{2}\hat{\boldsymbol{\lambda}}^\top\boldsymbol{u}_{ij}\boldsymbol{u}_{ij}^\top\hat{\boldsymbol{\lambda}}\right)+o_p(1)\\
&=2n\left(\hat{\boldsymbol{\lambda}}^\top\boldsymbol{U}-\frac{1}{2}\hat{\boldsymbol{\lambda}}^\top\boldsymbol{D}\hat{\boldsymbol{\lambda}}\right)+o_p(1)\\
&=n\boldsymbol{U}^\top\boldsymbol{D}^{-1}\boldsymbol{U}+o_p(1)\\
&=n\scalemath{0.9}{\begin{pmatrix}
0 & \thetaH-\theta
\end{pmatrix}
\begin{pmatrix}
d^{(11)} & d^{(12)}\\
d^{(21)} & d^{(22)}
\end{pmatrix}
\begin{pmatrix}
0\\ \thetaH-\theta
\end{pmatrix}+o_p(1)}\\
&=nd^{(22)}(\thetaH-\theta)^2+o_p(1)
\; ,
\end{aligned}
\label{ratio}
\end{equation}
where $d^{(22)}$ is the $(2,2)$-th element of $\boldsymbol{D}^{-1}$. 
The second equality in (\ref{ratio}) holds since  $\hat{\boldsymbol{\lambda}}^\top\boldsymbol{u}_{ij}=o_p(1)$ uniformly over $i, j$. 
A standard calculation yields that
\begin{equation}
\begin{aligned}
    d^{(22)}=\Big\{\theta^2+2\Big[\sum_{j\in \bst}\tilde{a}_{1j}Y_{1j}^2+\sum_{j\in \bsc}\tilde{a}_{0j}Y_{0j}^2-\thetaH\theta\Big]- (\mutH+\mucH)^2\Big\}^{-1},
\end{aligned}
\notag
\end{equation}
which converges in probability to
\[    
    d_0^{(22)}=\big\{2[\operatorname{E}(Y_{1j}^2+Y_{0j}^2)-(\mu_1^0)^2-(\mu_0^0)^2]\big\}^{-1}\, ,
\]
under the assumed propensity score model, with $\theta=\theta^0=\mu_1^0-\mu_0^0$. 

By using the estimating equation techniques and the Taylor expansion, under the assumed propensity score model, we get
\[\scalemath{0.81}{
\thetaH-\theta^0=\frac{1}{n}\sum_{j=1}^n\left[\frac{\Ri}{\tau_j^0}\left(\Yti-\mu^0_1\right)-\frac{1-\Ri}{1-\tau_j^0}\left(\Yci-\mu^0_0\right)-(\boldsymbol{J}-\boldsymbol{G})\boldsymbol{C}^{-1}\bz(\Ri-\tau_j^0)\right]+o_p(n^{-1/2})}\, ,
\]
where $\tau_j^0=\tau(\bz;\boldsymbol{\alpha}^0)$ is the propensity score under the true propensity score model with $\boldsymbol{\alpha}^0$ representing the true value of $\balpha$, $\boldsymbol{C}=-\operatorname{E}[\tau_j^0(1-\tau_j^0)\bz\bz^\top]$, $\boldsymbol{J}=-\operatorname{E}[\Ri(Y_{1j}-\mu_1^0)(1-\tau_j^0)\bz^\top/\tau_j^0]$, and $\boldsymbol{G}=\operatorname{E}[(1-\Ri)(Y_{0j}-\mu_0^0)(1-\tau_j^0)^{-1}\tau_j^0\bz^\top]$.

It follows that, under the regularity conditions {\bf R2} and {\bf R6}, we have
\[
\sqrt{n}\left(\thetaH-\theta^0\right)\stackrel{d}{\rightarrow}\operatorname{N}(0, V)\;\;{\rm as } \;\; n\rightarrow \infty\, ,
\]
where $V=n\operatorname{Var}(\thetaH)$ and $\stackrel{d}{\rightarrow}$ denotes convergence in distribution.
By Slutsky's theorem, we can conclude that under the assumed propensity score model, 
\[
    -2r_{\mbox{\tiny PEL}}(\theta^0)/\hat{c}\stackrel{d}{\rightarrow}\chi^2_1\, ,
\]
where $\hat{c}=n\hat{d}_0^{(22)}\operatorname{var}(\thetaH)$.
This completes the proof.
\end{proof}

\subsection{Proof of Theorem \ref{mutPCch2}}
\label{mutPCch2proof}

\begin{proof}
We first derive an expression for the maximizer $\hat{\boldsymbol{p}}_1$. The original constrained optimization problem is equivalent to maximizing the pseudo-empirical likelihood function 
$nw_1\sum_{j\in \bs_1}\tilde{a}_{1j}\log(p_{1j})
$
under the normalization constraint
$
\sum_{j\in \bs_1} p_{1j}=1
$
and the outcome regression model constraint
$
\sum_{j \in \bs_{1}} p_{1j}\hat{m}_{1j}=\mtbar \, .
$ 
Under the normalization constraint, we can rewrite the model-calibration constraint as $\sum_{j \in \bst}p_{1j}\hat{u}_{1j}=0$. Let
\[
    \mathrm{L}(\mbox{\boldmath{$p$}}_1)=nw_1 \sum_{j \in \bst} \tilde{a}_{1j} \log \left(p_{1j}\right)-\lambda_{11}\left(\sum_{j \in \bst} p_{1j}-1\right)-\lambda_{12}\left(\sum_{j \in \bst} p_{1j}\hat{u}_{1j}\right)\, ,
\]
where $\lambda_{11}$ and $\lambda_{12}$ are the Lagrange multipliers. 
Differentiating $\mathrm{L}(\mbox{\boldmath{$p$}}_1)$ with respect to $p_{1j}$ and setting these equations to be zero leads to
 \begin{equation}
    \partial \mathrm{L}(\mbox{\boldmath{$p$}}_{1}) / \partial p_{1j} = nw_1 \tilde{a}_{1j} / p_{1j}-\lambda_{11}-\lambda_{12} \hat{u}_{1j}=0 \,.
    \label{diff1}   
 \end{equation}
It follows that $nw_1\tilde{a}_{1j}-\hat{\lambda}_{11}\hat{p}_{1j}-\hat{\lambda}_{12}\hat{u}_{1j}\hat{p}_{1j}=0$. By summing over all the subjects in $\bst$, we get $\hat{\lambda}_{11}=nw_1$. Putting this back to (\ref{diff1}) leads to
\[
    \hat{p}_{1j}=\frac{nw_1\tilde{a}_{1j}}{\hat{\lambda}_{11}+\hat{\lambda}_{12}\hat{u}_{1j}}=\frac{1}{\hat{\lambda}_{11}}\frac{nw_1\tilde{a}_{1j}}{1+\hat{\lambda}_{1} \hat{u}_{1j}}=\frac{\tilde{a}_{1j}}{1+\hat{\lambda}_{1} \hat{u}_{1j}}\, ,
\]
where $\hat{\lambda}_{1}=\hat{\lambda}_{12}/\hat{\lambda}_{11}$ satisfies the model-calibration constraint $\sum_{j \in \bst}p_{1j}\hat{u}_{1j}=0$, i.e., 
\begin{equation}
\sum_{j\in \bst} \frac{\tilde{a}_{1j}\hat{u}_{1j}}{1+\hat{\lambda}_{1} \hat{u}_{1j}}=0\, . 
\label{reformation }
\end{equation}

\medskip

\noindent
{\em (1) The propensity score model is correctly specified.}

\medskip

The equation (\ref{reformation }) can be rewritten as 
\[
 \sum_{j\in \bst} \frac{\tilde{a}_{1j}\hat{u}_{1j}(1+\hat{\lambda}_{1} \hat{u}_{1j}-\hat{\lambda}_{1} \hat{u}_{1j})}{1+\hat{\lambda}_{1} \hat{u}_{1j}}=0\, ,   
\]
which implies
\begin{equation}
    \hat{\lambda}_{1}=\left(\sum_{j \in \bst} \frac{\tilde{a}_{1j} \hat{u}_{1j}^{2}}{1+\hat{\lambda}_{1} \hat{u}_{1j}}\right)^{-1}\left(\sum_{j \in \bst} \tilde{a}_{1j} \hat{u}_{1j}\right)\, .
\label{reformulated}
\end{equation}
On the other hand, it follows from (\ref{reformation }) that
\[
    \left|\sum_{j \in \bst} \tilde{a}_{1j} \hat{u}_{1j}\right|=\left|\hat{\lambda}_{1}\right|\left|\sum_{j \in \bst} \frac{\tilde{a}_{1j} \hat{u}_{1j}^{2}}{1+\hat{\lambda}_{1} \hat{u}_{1j}}\right|\geq\frac{\left|\hat{\lambda}_{1}\right|}{1+\left|\hat{\lambda}_{1}\right|u_1^*}\sum_{j \in \bst}\tilde{a}_{1j} \hat{u}_{1j}^{2}\, ,
\]
where $u_1^*=\max_{j\in \bst}|\hat{u}_{1j}|$. 
Now, we demonstrate that $u_1^*$ is of order $o_p(n^{1/2})$. 
By the Taylor expansion of $\hat{u}_{1j}$ on $\hat{\boldsymbol{\beta}}_1$ around $\boldsymbol{\beta}^*_1$, under the regularity condition {\bf R5}, we have
\[
    \hat{u}_{1j}=\hat{m}_{1j}-\bar{\hat{m}}_{1}=(m_{1j}^*-\bar{m}_1^*)+ \left(\frac{\partial (m_{1j}-\bar{m}_1)}{ \partial \boldsymbol{\beta}_1}   \Big | _{\boldsymbol{\beta}_1=\boldsymbol{\beta}_1^*}\right)\left(\boldsymbol{\hat{\beta}}_1-\boldsymbol{\beta}^*_1\right) +o_p(1)\, ,
\]
where $\bar{m}_1=n^{-1}\sum_{l=1}^n m_{1l}$ and $m_{1j}^*=m_1(\bx,\boldsymbol{\beta}^*_1)$. 
Under the regularity conditions {\bf R1} and {\bf R3}-{\bf R5}, it is true that $u_1^*=o_p(n^{1/2})$.  
Similarly, by Taylor expansions, we get that $ |\sum_{j \in \bst} \tilde{a}_{1j} \hat{u}_{1j}|$ is of the order $O_p(n^{-1/2})$, and $\sum_{j \in \bst}\tilde{a}_{1j} \hat{u}_{1j}^{2}$ is of order $O_p(1)$ under the regularity conditions {\bf R1} and {\bf R3}-{\bf R4}. 
Thus, we derive the order of $\hat{\lambda}_{1}$ to be $O_p(n^{-1/2})$, which further implies $\hat{\lambda}_{1} \hat{u}_{1j}=o_p(1)$ uniformly over all $j\in \bst$. 
The above asymptotic analysis leads to 
\[
    \hat{\lambda}_{1}=\left(\sum_{j \in \bst} \tilde{a}_{1j} \hat{u}_{1j}^{2}\right)^{-1}\left(\sum_{j \in \bst} \tilde{a}_{1j} \hat{u}_{1j}\right)+o_{p}\left(n^{-1 / 2}\right)\, .
\]

Under regularity conditions {\bf R1}-{\bf R3}, the maximum PEL estimator for $\mu_1$ has an asymptotic expansion given by 
\begin{equation}
    \begin{aligned} 
    \hat{\mu}_{1\mbox{\tiny MCP}} &=\sum_{j \in \bst} \hat{p}_{1j} Y_{1j}=\sum_{j \in \bst} \frac{\tilde{a}_{1j}}{1+\hat{\lambda}_{1} \hat{u}_{1j}} Y_{1j} \\ & = \sum_{j \in \bst} \tilde{a}_{1j}\left(1-\hat{\lambda}_{1} \hat{u}_{1j}\right) Y_{1j}+o_p(n^{-1/2}) \\ &=\sum_{j \in \bst} \tilde{a}_{1j} Y_{1j}+\hat{B}\left\{\frac{1}{n}\sum_{j\in\bs}\hat{m}_{1j}-\sum_{j \in \bst} \tilde{a}_{1j} \hat{m}_{1j}\right\}+o_{p}\left(n^{-1 / 2}\right)\, , \end{aligned}
    \notag
\end{equation}
where $\hat{B}=\{\sum_{j \in \bst} \tilde{a}_{1j}\hat{u}_{1j}^{2}\}^{-1}\{\sum_{j \in \bst} \tilde{a}_{1j}\hat{u}_{1j}Y_{1j}\} $. 
The expression for $\hat{\mu}_{1\mbox{\tiny MCP}}$ shows that
\[
\hat{\mu}_{1\mbox{\tiny MCP}}=\mutH-\hat{B}\left(\sbstj\tilde{a}_{1j}\hat{u}_{1j}\right)+o_p(n^{-1/2})=\mu_1^0+o_p(1)\; ,
\]
where $\hat{B}=\left\{\operatorname{Var}\left(m_{1j}^*\right)\right\}^{-1}\operatorname{Cov}\left(m_{1j}^*,\Yti\right)+o_p(1)=O_p(1)$.

\medskip

\noindent
{\em (2) The outcome regression model is correctly specified.}

\medskip

The maximum PEL estimator can be expressed as
\[
\hat{\mu}_{1\mbox{\tiny MCP}}=\sbstj\hat{p}_{1j}\Yti=\sbstj\hat{p}_{1j}\left(\Yti-\hat{m}_{1j}\right)+\sbstj\hat{p}_{1j}\hat{m}_{1j}\, .
\]
Applying the Taylor expansion to the first term of the right hand side of the above equation gives
\begin{equation}
\begin{aligned}
\sbstj\hat{p}_{1j}\left(\Yti-\hat{m}_{1j}\right)=&\sum_{j=1}^n\frac{\Ri\tilde{a}_{1j}}{1+\hat{\lambda}_{1}\hat{u}_{1j}}\left(\Yti-\hat{m}_{1j}\right)\\
=&\frac{1}{n}\sum_{j=1}^n\frac{\Ri\left(\Yti-{m}_{1j}^*\right)}{\tau_j^*\left(1+\lambda_1^*u_{1j}^*\right)}\Big/\frac{1}{n}\sum_{j=1}^n\frac{\Ri}{\tau_j^*}+o_p(1)\; , 
\end{aligned}
 \notag   
\end{equation}
which converges in probability to 
\[
\operatorname{E}\left(\frac{\Ri\left(\Yti-m_{1j}^*\right)}{\tau_j^*\left(1+\lambda_1^*u_{1j}^*\right)}\right)\Big/\operatorname{E}\left(\frac{\Ri}{\tau_j^*}\right)\; ,
\]
under regularity conditions {\bf R1}-{\bf R4}, where $\tau_j^*=\tau(\bz,\boldsymbol{\alpha}^*)$, $\lambda_1^*=\lambda(\boldsymbol{\alpha}^*,\boldsymbol{\beta}_1^*)$, and $u_{1j}^*=m_{1j}^*-\bar{m}_1^*$.
When the outcome regression model is correctly specified, by the law of total expectation conditional on $\bx$, we get $\operatorname{E}[\{\tau_j^*(1+\lambda_1^*u_{1j}^*)\}^{-1}\{\Ri(\Yti-m_{1j}^*)\}]=0$. 
Therefore, $\hat{\mu}_{1\mbox{\tiny MCP}}=\sbstj\hat{p}_{1j}\hat{m}_{1j}+o_p(1)$. 

On the other hand, from the model-calibration constraint, we obtain that 
\[
\sbstj\hat{p}_{1j}\hat{m}_{1j}=\frac{1}{n}\sbsj\hat{m}_{1j}\; .
\]
Under the regularity condition {\bf R4}, applying the Taylor expansion immediately leads to 
\[
\sbstj\hat{p}_{1j}\hat{m}_{1j}=\frac{1}{n}\sbsj m_{1j}^*+o_p(1)\stackrel{P}{\rightarrow}\operatorname{E}\left(m_{1j}^*\right)=\mu_1^0\; ,
\]
where $\stackrel{P}{\rightarrow}$ denotes convergence in probability. Thus, when the outcome regression model is correct, $\hat{\mu}_{1\mbox{\tiny MCP}}$ is a consistent estimator of $\mu_1$.    
\end{proof}

\subsection{Proof of Theorem \ref{normalPoint}}
\label{normalPointproof}

\begin{proof}
We first note that  $\hat{p}_{ij}(\theta)=\aij/\{1+\hat{\boldsymbol{\lambda}}^\top\gij\}$, 
where $\hat{\boldsymbol{\lambda}}$ is the solution to 
\[
\sumzo w_i \sbsji \frac{\aij \gij}{1+\boldsymbol{\lambda}^\top \gij}=\boldsymbol{0}\, .
\]
We also note that $\hat{\boldsymbol{\lambda}}$ is a function of $\theta$, so it can be denoted as $\hat{\boldsymbol{\lambda}}(\theta)$.
If the propensity score model is correctly specified, for $\theta=\theta^0+O_p(n^{-1/2})$ and under regularity conditions {\bf R1}-{\bf R5}, we have 

(i) $\sumzo w_i\sbsji\aij\gij=O_p(n^{-1/2})$; 

(ii) $\max_{i,j}\left\|\gij\right\|=o_p(n^{1/2})$; 

(iii) $\sumzo w_i \sbsji \aij \gij \gij^\top=O_p(1)$.

\noindent
The results (i)-(iii) altogether imply that $\hat{\boldsymbol{\lambda}}(\theta)=O_p(n^{-1/2})$. Let
\[
Q_{n 1}(\theta, \boldsymbol{\lambda})=\sumzo w_i \sbsji \frac{\aij \gij}{1+\boldsymbol{\lambda}^\top \gij}\]
and
\[
    Q_{n 2}(\theta, \boldsymbol{\lambda})=\sumzo w_i \sbsji \frac{\aij}{1+\boldsymbol{\lambda}^{\top} \gij}\left\{\frac{\partial \gij}{\partial \theta^{\top}}\right\}^{\top} \boldsymbol{\lambda}\; .
\]
The estimators $\thetaPC$ and $\lambdaPC=\hat{\boldsymbol{\lambda}}(\thetaPC)$ satisfy that $Q_{n 1}(\thetaPC, \lambdaPC)=0$ and $Q_{n 2}(\thetaPC, \lambdaPC)=0$. Applying Taylor expansions to $Q_{n 1}(\thetaPC, \lambdaPC)$ and $Q_{n 2}(
\thetaPC, \lambdaPC)$ at $(\theta^0, \boldsymbol{0})$ yields 
\begin{equation}
\scalemath{0.95}{
    \begin{aligned}
&Q_{n 1}\left(\theta^0, \boldsymbol{0}\right)+\frac{\partial Q_{n 1}\left(\theta^0, \boldsymbol{0}\right)}{\partial \theta}\left(\thetaPC-\theta^0\right)+\frac{\partial Q_{n 1}\left(\theta^0, \boldsymbol{0}\right)}{\partial \boldsymbol{\lambda}}\left(\lambdaPC-\boldsymbol{0}\right)+o_{p}\left(\sigma_{n}\right)=\boldsymbol{0}\, ,\\
&Q_{n 2}\left(\theta^0, \boldsymbol{0}\right)+\frac{\partial Q_{n 2}\left(\theta^0, \boldsymbol{0}\right)}{\partial \theta}\left(\thetaPC-\theta^0\right)+\frac{\partial Q_{n 2}\left(\theta^0, \boldsymbol{0}\right)}{\partial \boldsymbol{\lambda}}\left(\lambdaPC-\boldsymbol{0}\right)+o_{p}\left(\sigma_{n}\right)=\boldsymbol{0}\, ,
    \end{aligned}
\notag}
\end{equation}
where $\sigma_{n}=\|\thetaPC-\theta^0\|+\|\lambdaPC\|=O_p(n^{-1/2})$. Thus, a standard calculation leads to
\[
\left(\begin{array}{c}-Q_{n 1}\left(\theta^0, \boldsymbol{0}\right)+o_{p}\left(n^{-1/2}\right) \\ o_{p}\left(n^{-1/2}\right)\end{array}\right)
=\boldsymbol{S}_{n1}
\left(\begin{array}{c}\lambdaPC \\ \thetaPC-\theta^0\end{array}\right)\, ,
\]
where 
\[
\boldsymbol{S}_{n1}\stackrel{P}{\rightarrow}\left(\begin{array}{cc}-\Wone & \Gam \\ \Gam^{\top} & 0\end{array}\right)\, .
\]
By noting that $Q_{n 1}\left(\theta^0, \boldsymbol{0}\right)=O_p(n^{-1/2})$, the above equation implies that 
\[
\left(\begin{array}{c}\lambdaPC \\ \thetaPC-\theta^0\end{array}\right)=\left(\begin{array}{cc}-\Pone & \Sigone\Wone^{-1}\Gam \\ \Sigone\Gam^{\top}\Wone^{-1} & \Sigone\end{array}\right)
\left(\begin{array}{c}-Q_{n 1}\left(\theta^0, \boldsymbol{0}\right) \\ 0\end{array}\right)+o_p(n^{-1/2})\; ,
\]
where $\Sigone=(\Gam^\top\Wone^{-1}\Gam)^{-1}$ and $\Pone=\Wone^{-1}-\Sigone\Wone^{-1}\Gam\Gam^\top\Wone^{-1}$.

On the other hand, we have the expression for $Q_{n 1}\left(\theta^0, \boldsymbol{0}\right)$ that
\[
Q_{n 1}\left(\theta^0, \boldsymbol{0}\right)=\frac{1}{n}\sbsj\boldsymbol{h}_j+o_p(n^{-1/2})\; ,
\]
where $\boldsymbol{h}_j$ is defined in Section \ref{PELRBCI}. Note that $\operatorname{E}(\boldsymbol{h}_j)=\boldsymbol{0}$ and $\operatorname{Var}(\boldsymbol{h}_j)<\infty$ under regularity conditions {\bf R2}-{\bf R3} and {\bf R6}.
It follows that $\sqrt{n}Q_{n 1}\left(\theta^0, \boldsymbol{0}\right)\stackrel{d}{\rightarrow}\operatorname{MVN}\left(\boldsymbol{0}, \boldsymbol{\Omega}\right)$. Therefore, 
\[
\sqrt{n}\left(\thetaPC-\theta^0\right)=-\Sigone\Gam^\top\Wone^{-1}{\sqrt{n}Q_{n 1}\left(\theta^0, \boldsymbol{0}\right)}+o_p(1)\stackrel{d}{\rightarrow}\operatorname{N}\left(0,V_1\right)\; ,
\]
where $V_1=\Sigone^2\Gam^\top\Wone^{-1}\boldsymbol{\Omega}\Wone^{-1}\Gam$.
\end{proof}

\subsection{Proof of Theorem  \ref{PELWMCCThm}}
\label{PELWMCCThmproof}

\begin{proof}
The Lagrange multiplier $\boldsymbol{\lambda}$ is a function of $\theta$, i.e., $\boldsymbol{\lambda}=\boldsymbol{\lambda}(\theta)$, which is the solution to
\begin{equation}
\sumzo w_i\sbsji\frac{\aij\gij}{1+\boldsymbol{\lambda}^\top\gij}=\boldsymbol{0}\; .   
\label{funlambda}
\end{equation}
For $\theta=\theta^0+O_p(n^{-1/2})$, we have $\boldsymbol{\lambda}=O_p(n^{-1/2})$ as argued in Section \ref{normalPointproof}. Taking the Taylor expansion of (\ref{funlambda}) around $\boldsymbol{\lambda}=\boldsymbol{0}$ gives 
\[
\boldsymbol{0}=\sumzo w_i\sbsji \aij\gij-\sumzo w_i\sbsji\aij\gij\gij^\top\boldsymbol{\lambda}+o_p(n^{-1/2})\; ,
\]
which implies that, at $\theta=\theta^0$,
\begin{equation}
\begin{aligned}
\boldsymbol{\lambda}&=\left[\sumzo w_i\sbsji\aij\gij\gij^\top\right]^{-1}\left(\sumzo w_i\sbsji\aij\gij\right)+o_p(n^{-1/2})\\
&=\Wone^{-1}Q_{n 1}\left(\theta^0, \boldsymbol{0}\right)+o_p(n^{-1/2})\; .
\end{aligned}
\notag
\end{equation}
which further leads to 
\begin{equation}
\scalemath{0.88}{
    \begin{aligned}
    -2\ell_{\mbox{\tiny PEL}}(\hat{\mbox{\boldmath{$p$}}}_1(\theta), \hat{\mbox{\boldmath{$p$}}}_0(\theta))&=nA+2n\sumzo w_i \sbsji\aij\log\left\{1+\boldsymbol{\lambda}^\top\gij\right\}\\
    &=nA+2n\sumzo w_i\sbsji\aij\left\{\boldsymbol{\lambda}^\top\gij-\frac{1}{2}\boldsymbol{\lambda}^\top\gij\gij^\top\boldsymbol{\lambda}\right\}+o_p(1)\\
    &=nA+nQ_{n 1}\left(\theta^0, \boldsymbol{0}\right)^\top\Wone^{-1}Q_{n 1}\left(\theta^0, \boldsymbol{0}\right)+o_p(1)\; ,
    \end{aligned}
    \notag}
\end{equation}
where $A=-2\left\{\sumzo w_i\sbsji\aij\log\aij\right\}$.

It is shown in Theorem \ref{normalPoint} that $\lambdaPC=\boldsymbol{P}_1Q_{n 1}\left(\theta^0, \boldsymbol{0}\right)+o_p(n^{-1/2})$, therefore, 
\begin{equation}\scalemath{0.93}{
    \begin{aligned}
-2\ell_{\mbox{\tiny PEL}}\left(\hat{\mbox{\boldmath{$p$}}}_1\left(\thetaPC\right), \hat{\mbox{\boldmath{$p$}}}_0\left(\thetaPC\right)\right)=&nA+2n\sumzo w_i \sbsji\aij\log\left\{1+\lambdaPC^\top\gijhat\right\}\\
=&nA+nQ_{n 1}\left(\theta^0, \boldsymbol{0}\right)^\top\boldsymbol{P}_1\Wone\boldsymbol{P}_1Q_{n 1}\left(\theta^0, \boldsymbol{0}\right)+o_p(1)\\
=&nA+nQ_{n 1}\left(\theta^0, \boldsymbol{0}\right)^\top\boldsymbol{P}_1Q_{n 1}\left(\theta^0, \boldsymbol{0}\right)+o_p(1)\, .
    \end{aligned}
    \notag}
\end{equation}

We can conclude that, when $\theta=\theta^0$,
\[
-2r_{\mbox{\tiny PEL}}(\theta)=n\Sigone Q_{n 1}\left(\theta^0, \boldsymbol{0}\right)^\top\Wone^{-1}\Gam\Gam^\top\Wone^{-1}Q_{n 1}\left(\theta^0, \boldsymbol{0}\right)+o_p(1)\stackrel{d}{\rightarrow}\boldsymbol{Q}^\top\boldsymbol{M}\boldsymbol{Q}\; ,
\]
where $\boldsymbol{M}=\Sigone\Ome^{1/2}\Wone^{-1}\Gam\Gam^\top\Wone^{-1}\Ome^{1/2}$, $\boldsymbol{Q}\sim \operatorname{MVN}(\boldsymbol{0}, \boldsymbol{I}_4)$, and $\boldsymbol{I}_4$ denotes the $4\times 4 $ identity matrix. This implies that $-2r_{\mbox{\tiny PEL}}(\theta^0)\stackrel{d}{\rightarrow}\delta\chi^2_1$, where $\delta$ is the non-zero eigenvalue of $\boldsymbol{M}$.
\end{proof}

\subsection{Justification of the Bootstrap Procedure}
\label{validitybootstrap}

\noindent
{\em (1) The propensity score model is correctly specified.}

\medskip

First, we can argue that each bootstrap sample leads to a similar asymptotic expansion in the form of 
\[\scalemath{0.8}{
\left(\begin{array}{c}\lambdaPCB \\ \thetaPCB-\thetaPC\end{array}\right)=\left(\begin{array}{cc}-\Pone^{[b]} & \left(\Wone^{[b]}\right)^{-1}\Gam\Sigone^{[b]} \\ \Sigone^{[b]}\Gam^{\top}\left(\Wone^{[b]}\right)^{-1} & \Sigone^{[b]}\end{array}\right)
\left(\begin{array}{c}-Q^{[b]}_{n 1}\left(\thetaPC, \boldsymbol{0}\right) \\ 0\end{array}\right)+o_p(n^{-1/2})\; ,}
\]
where $\thetaPCB$ is the bootstrap version of the point estimator obtained via maximizing the joint pseudo-empirical likelihood function $\ell^{[b]}_{\mbox{\tiny PEL}}\left(\boldsymbol{p}_{1}, \boldsymbol{p}_{0}\right)$ subject to the normalization and model-calibration constraints, and $\lambdaPCB$ is the corresponding value for the parameter $\boldsymbol{\lambda}$. 
Let $\Wone^{[b]}$ denote the limit of $\sumzo w_i\sum_{j\in\bs_i^{[b]}}\tilde{a}_{ij}^{[b]}\boldsymbol{g}_{ij}^{[b]}$\\$(\thetaPC)\gijhatstar^\top$.
We have $\Sigone^{[b]}=(\Gam^\top(\Wone^{[b]})^{-1}\Gam)^{-1}$ and $\Pone^{[b]}=(\Wone^{[b]})^{-1}-\Sigone^{[b]}(\Wone^{[b]})^{-1}\Gam\Gam^\top(\Wone^{[b]})^{-1}$. These quantities are just the bootstrap versions of the $\Sigone$, $\Pone$ and $\Wone$. The bootstrap version of $Q_{n 1}\left(\theta, \boldsymbol{\lambda}\right)$ is
\[
Q^{[b]}_{n 1}(\theta, \boldsymbol{\lambda})=\sumzo w_i \sum_{j\in\bs_i^{[b]}} \frac{\aijs \gijs}{1+\boldsymbol{\lambda}^\top \gijs}\,,
\]
and $Q^{[b]}_{n 1}(\thetaPC, \boldsymbol{0})$ is also asymptotically normally distributed. 

Second, we can argue that $(\thetaPC, \boldsymbol{\lambda}_B^{[b]})$ satisfies that $Q^{[b]}_{n 1}(\thetaPC, \boldsymbol{\lambda}_B^{[b]})=0$, where $\boldsymbol{\lambda}_B^{[b]}=\boldsymbol{\lambda}(\thetaPC)$ based on $Q^{[b]}_{n 1}(\theta, \boldsymbol{\lambda})$. 
Applying the Taylor expansion to $Q^{[b]}_{n 1}(\thetaPC, \boldsymbol{\lambda}_B^{[b]})=\boldsymbol{0}$ at $\boldsymbol{\lambda}_B^{[b]}=\boldsymbol{0}$ gives that $\boldsymbol{\lambda}_B^{[b]}=(\Wone^{[b]})^{-1}Q^{[b]}_{n 1}(\thetaPC, \boldsymbol{0})+o_p(n^{-1/2})$. This yields the expression
\[\scalemath{0.87}{
2\ell^{[b]}_{\mbox{\tiny PEL}}\left(\hat{\boldsymbol{p}}_{1}\left(\thetaPC\right), \hat{\boldsymbol{p}}_{0}\left(\thetaPC\right)\right)=nA^{[b]}-nQ^{[b]}_{n 1}\left(\thetaPC, \boldsymbol{0}\right)^\top(\Wone^{[b]})^{-1}Q^{[b]}_{n 1}\left(\thetaPC, \boldsymbol{0}\right)+o_p(1)}\, ,
\]
where $A^{[b]}=\sumzo\sum_{j\in\bs_i^{[b]}}\aijs\log\aijs$. On the other hand, we compute $\ell^{[b]}_{\mbox{\tiny PEL}}(\boldsymbol{p}_{1}$\\$(\theta), \boldsymbol{p}_{0}\left(\theta\right))$ at $\theta=\thetaPCB$ to obtain that 
\[\scalemath{0.91}{
2\ell^{[b]}_{\mbox{\tiny PEL}}\left(\hat{\boldsymbol{p}}_{1}\left(\thetaPCB\right), \hat{\boldsymbol{p}}_{0}\left(\thetaPCB\right)\right)=nA^{[b]}-nQ^{[b]}_{n 1}\left(\thetaPC, \boldsymbol{0}\right)^\top\Pone^{[b]} Q^{[b]}_{n 1}\left(\thetaPC, \boldsymbol{0}\right)+o_p(1)}\, .
\]

Finally, the above results lead to 
\begin{equation}\scalemath{0.95}{
    \begin{aligned}
 -2r^{[b]}_{\mbox{\tiny PEL}}(\thetaPC)&=n\Sigone^{[b]} Q^{[b]}_{n 1}\left(\thetaPC, \boldsymbol{0}\right)^\top(\Wone^{[b]})^{-1}\Gam\Gam^\top(\Wone^{[b]})^{-1}Q^{[b]}_{n 1}\left(\thetaPC, \boldsymbol{0}\right)+o_p(1)\\
 &\stackrel{d}{\rightarrow}\boldsymbol{Q}^{\top}\boldsymbol{M}^{[b]}\boldsymbol{Q}\; ,          \end{aligned}
 \notag}
\end{equation}
where $\boldsymbol{M}^{[b]}=\Sigone^{[b]}(\Ome^{[b]})^{1/2}(\Wone^{[b]})^{-1}\Gam\Gam^\top(\Wone^{[b]})^{-1}(\Ome^{[b]})^{1/2}$ and $\boldsymbol{Q}\sim \operatorname{MVN}(\boldsymbol{0}, \boldsymbol{I}_4)$. Moreover, $\Ome^{[b]}$ is the asymptotic variance matrix of $Q^{[b]}_{n 1}(\thetaPC, \boldsymbol{0})$. This implies that $-2r^{[b]}_{\mbox{\tiny PEL}}(\theta)\stackrel{d}{\rightarrow}\delta^{[b]}\chi^2_1$ if $\theta=\thetaPC$, where $\delta^{[b]}$ is the non-zero eigenvalue of $\boldsymbol{M}^{[b]}$. The matrix $\boldsymbol{M}^{[b]}$ is the bootstrap sample version of the matrix $\boldsymbol{M}$ where we obtain $\delta$. Therefore, $\delta^{[b]}$ converges in probability to $\delta$. 

\medskip

\noindent
{\em (2) The outcome regression models are correctly specified.}

\medskip

We justify the bootstrap procedure in this case by arguing that every component of the pseudo-empirical likelihood ratio function from the bootstrap sample converges in probability to the same limit as that from the original sample. Thus, the pseudo-empirical likelihood ratios should have the same limiting distribution.

The pseudo-empirical likelihood ratio for the original sample at $\theta=\theta^0$ is given by  
\begin{equation}\scalemath{0.84}{
\begin{aligned}
r_{\mbox{\tiny PEL}}\left(\theta^0\right)&=\ell_{\mbox{\tiny PEL}}\left(\hat{\boldsymbol{p}}_{1}\left(\theta^0\right), \hat{\boldsymbol{p}}_{0}\left(\theta^0\right)\right)-\ell_{\mbox{\tiny PEL}}\left(\hat{\boldsymbol{p}}_{1}\left(\thetaPC\right), \hat{\boldsymbol{p}}_{1}\left(\thetaPC\right)\right)\\
&=n\sumzo w_i \sbsji\tilde{a}_{ij}\log{\left(1+\lambdaPC^\top\boldsymbol{g}_{ij}\left(\thetaPC\right)\right)}-n\sumzo w_i \sbsji\tilde{a}_{ij}\log{\left(1+\boldsymbol{\lambda}^\top\boldsymbol{g}_{ij}\left(\theta^0\right)\right)}\; ,
\end{aligned}
\label{PELRoriginal}}
\end{equation}
where $\lambdaPC$ is the solution to 
\begin{equation}
\sumzo w_i \sbsji \frac{\aij \boldsymbol{g}_{ij}\left(\thetaPC\right)}{1+\boldsymbol{\lambda}^\top \boldsymbol{g}_{ij}\left(\thetaPC\right)}=\boldsymbol{0}\,,
\label{lamequation2}   
\end{equation}
and $\boldsymbol{\lambda}$ is the solution to
\begin{equation}
\sumzo w_i \sbsji \frac{\aij \gijzero}{1+\boldsymbol{\lambda}^\top \gijzero}=\boldsymbol{0}\, .
\label{lamequationnull}   
\end{equation}
On the other hand, the pseudo-empirical likelihood ratio for the bootstrap sample, at $\theta=\thetaPC$, is
\begin{equation}\scalemath{0.75}{
    \begin{aligned}
    r^{[b]}_{\mbox{\tiny PEL}}\left(\thetaPC\right)&=\ell^{[b]}_{\mbox{\tiny PEL}}\left(\hat{\mbox{\boldmath{$p$}}}_1\left(\thetaPC\right), \hat{\mbox{\boldmath{$p$}}}_0\left(\thetaPC\right)\right)-\ell^{[b]}_{\mbox{\tiny PEL}}\left(\hat{\mbox{\boldmath{$p$}}}_1\left(\thetaPCB\right), \hat{\mbox{\boldmath{$p$}}}_0\left(\thetaPCB\right)\right)\\
    &=n\sumzo w_i \sum_{j\in\bs^{[b]}_i}\tilde{a}^{[b]}_{ij}\log{\left(1+\hat{\boldsymbol{\lambda}}_{\mbox{\tiny MCP}}^{[b]\top}\boldsymbol{g}_{ij}^{[b]}\left(\thetaPCB\right)\right)}-n\sumzo w_i \sum_{j\in\bs^{[b]}_i}\tilde{a}^{[b]}_{ij}\log{\left(1+\boldsymbol{\lambda}_B^{[b]\top}\boldsymbol{g}_{ij}^{[b]}\left(\thetaPC\right)\right)},
    \end{aligned}
    \label{PELRBoot}}
\end{equation}
where $\lambdaPCB$ is the solution to
\begin{equation}
\sumzo w_i \sum_{j\in\bs^{[b]}_i} \frac{\tilde{a}^{[b]}_{ij} \boldsymbol{g}_{i j}^{[b]}\left(\thetaPCB\right)}{1+\boldsymbol{\lambda}^\top \boldsymbol{g}_{i j}^{[b]}\left(\thetaPCB\right)}=\boldsymbol{0}\,, 
\label{lamequationBmax}   
\end{equation}
and $\boldsymbol{\lambda}_B^{[b]}$ is the solution to
\begin{equation}
\sumzo w_i \sum_{j\in\bs^{[b]}_i} \frac{\tilde{a}^{[b]}_{ij} \boldsymbol{g}_{i j}^{[b]}\left(\thetaPC\right)}{1+\boldsymbol{\lambda}^\top \boldsymbol{g}_{i j}^{[b]}\left(\thetaPC\right)}=\boldsymbol{0}\, .
\label{lamequationBnull}   
\end{equation}
We aim to compare (\ref{PELRoriginal}) with (\ref{PELRBoot}).

Note that $\tilde{a}_{ij}$, $i=0,1$ and $j\in\bs_i$, is calculated by $\hat{\tau}_{j}$ for $j\in\bs$. Note also that $\hat{\tau}_{j}=\tau(\boldsymbol{x}_j; \hat{\boldsymbol{\alpha}})$, and the estimating equation for $\hat{\boldsymbol{\alpha}}$ is defined as
\[
\frac{1}{n}\sum_{j\in\bs}\bz\left(\Ri-\tau\left(\boldsymbol{x}_j; {\boldsymbol{\alpha}}\right)\right)=\boldsymbol{0}\; .
\]
Suppose that $\hat{\boldsymbol{\alpha}}\stackrel{P}{\rightarrow}\boldsymbol{\alpha}^*$. For $\tilde{a}^{[b]}_{ij}$, analogously, $\hat{\tau}^{[b]}_{j}=\tau(\boldsymbol{x}_j; \hat{\boldsymbol{\alpha}}^{[b]})$, where $\hat{\boldsymbol{\alpha}}^{[b]}$ is the solution to
\[
\frac{1}{n}\sum_{j\in\bs^{[b]}}\bz(\Ri-\tau(\boldsymbol{x}_j; {\boldsymbol{\alpha}}))=\boldsymbol{0}\; .
\]
Even if the propensity score model is misspecified, the function $\tau$ preserves the same form in both the original and the bootstrap samples. Thus, we must have $\hat{\boldsymbol{\alpha}}^{[b]}\stackrel{P}{\rightarrow}\boldsymbol{\alpha}^*$. The correct specifications of the outcome regression models lead to that $\hat{\boldsymbol{\beta}}_i\stackrel{P}{\rightarrow}\boldsymbol{\beta}_i^0$. The estimating equations for $\hat{\boldsymbol{\beta}}_i$ and $\hat{\boldsymbol{\beta}}_i^{[b]}$ are respectively provided by
\[
\frac{1}{n_i}\sum_{j\in\bs_i}\bz(Y_{j}-m_i(\boldsymbol{x}_j; \boldsymbol{\beta}_i))=0     \;\;\;{\rm and} \;\;\;    
\frac{1}{n_i^{[b]}}\sum_{j\in\bs_i^{[b]}}\bz(Y_{j}-m_i(\boldsymbol{x}_j; \boldsymbol{\beta}_i))=0\; .
\]
Therefore, we get $\hat{\boldsymbol{\beta}}_i^{[b]}\stackrel{P}{\rightarrow}\boldsymbol{\beta}_i^0$.

Further, the pseudo-empirical likelihood function (\ref{jointPELboot}) and the constraints (\ref{constraintsB}) for the bootstrap sample are the bootstrap versions of the corresponding ones for the original sample ((\ref{jointPEL}), (\ref{jnorm}),  (\ref{para}), (\ref{jcal})) in terms of ($\bx$, $Y_j$, $\Ri$). So the estimator $\thetaPCB$, obtained through maximizing (\ref{jointPELboot}) subject to the first two constraints of (\ref{constraintsB}), is the corresponding bootstrap version of $\thetaPC$, calculated through maximizing (\ref{jointPEL}) subject to (\ref{jnorm}) and (\ref{jcal}), and it also convergences in probability to $\thetaPC$. Furthermore,  $\thetaPCB\stackrel{P}{\rightarrow}\theta^0$. 

Now, it remains to compare the different versions of $\boldsymbol{\lambda}$. Note that $\lambdaPC$ is the solution to (\ref{lamequation2}) and $\lambdaPCB$ is the solution to (\ref{lamequationBmax}), where (\ref{lamequationBmax}) is the bootstrap version of (\ref{lamequation2}) in terms of ($\bx$, $Y_j$, $\Ri$). Thus, we have $\lambdaPCB\stackrel{P}{\rightarrow}\lambdaPC$. Applying an analogous analysis on (\ref{lamequationnull}) and (\ref{lamequationBnull}) reveals that $\boldsymbol{\lambda}_B^{[b]}\stackrel{P}{\rightarrow}\boldsymbol{\lambda}$. 

Finally, we conclude that the pseudo-empirical likelihood ratio $r^{[b]}_{\mbox{\tiny PEL}}(\thetaPC)$ based on the bootstrap sample has asymptotically the same limiting distribution as the original one $r_{\mbox{\tiny PEL}}(\theta^0)$ even if the propensity score model is misspecified.  

\bigskip

\bibliographystyle{apalike}
\bibliography{EJS_PEL-2023-10-27_arXiv}

\end{document}